\documentclass[10pt,prd,nofootnotes]{revtex4}
\usepackage{eurosym}
\usepackage{amsfonts}
\usepackage{amsmath}
\usepackage{amssymb,epsf}
\usepackage{color}
\usepackage{graphicx}
\usepackage{epstopdf}
\usepackage{float}
\usepackage{comment}
\usepackage{caption}
\usepackage{subfig}
\usepackage{orcidlink}

\begin{document}

\title{Thermodynamics and optical aspects of ModMax black holes in higher order curvature gravity with quintessence dark energy
}
\author{Ahmad Al-Badawi \orcidlink{0000-0002-3127-3453}}
\email[Email: ]{ahmadbadawi@ahu.edu.jo
}
\affiliation{Department of Physics, Al-Hussein Bin Talal University, 71111,
Ma'an, Jordan.}
\author{Usman Zafar \orcidlink{0000-0001-9610-1081}} \email[Email: ]{s2471001@ipc.fukushima-u.ac.jp, zafarusman494@gmail.com}
\affiliation{Faculty of Symbiotic Systems Science, Fukushima University,
Fukushima 960-1296, Japan.}
\author{Abdul Jawad \orcidlink{0000-0001-5249-803X}} \email[Email: ]{abduljawad@cuilahore.edu.pk}
\affiliation{Department of Mathematics, COMSATS University
Islamabad, Lahore-Campus, Lahore-54000, Pakistan}
\author{Kazuharu Bamba \orcidlink{0000-0001-9720-8817}} \email[Email: ]{bamba@sss.fukushima-u.ac.jp}
\affiliation{Faculty of Symbiotic Systems Science, Fukushima University, Fukushima 960-1296, Japan}
\begin{abstract}
In this work, we derive an exact black hole solution in higher-order curvature gravity by coupling an electromagnetic sector formulated within the ModMax framework to a quintessence dark energy component. Focusing on purely electrically charged configurations, we analyze the thermodynamic and geothermodynamic properties of the solution to investigate its stability and phase structure. Within this sector, the ModMax theory effectively reduces Maxwell electrodynamics up to a rescaling of the electric charge, and thus the obtained solution corresponds to a consistent subset of the broader nonlinear theory. Using thermodynamic geometry, we examine microscopic interactions and phase transitions, showing that divergences in the thermodynamic curvature coincide with the vanishing of the heat capacity, confirming the consistency of the phase structure.
We further explore the optical properties of the black hole by studying null geodesics and determining the photon sphere and the corresponding shadow radius for different values of the quintessence state parameter $\omega$. Exact analytical expressions for the photon-sphere radius are derived, revealing that higher-order curvature corrections and quintessence significantly enhance the shadow size, whereas the electric charge has the opposite effect. Notably, quintessence is found to have a more pronounced impact on the shadow than the charge. These results highlight that dark energy and higher-order curvature corrections can yield potentially observable signatures in black hole shadows.
\end{abstract}

\maketitle

\section{Introduction}

Over the last few decades, the quest to understand the universe's accelerated expansion, which is often attributed
 to dark energy (DE), has driven extensive research into modified theories of gravity (MOG). The theoretical framework extends the foundation of general relativity (GR) and offers compelling substitutes for the $\Lambda$CDM model, which attributes the
 cosmological constant to describe DE. Although $\Lambda$CDM fits well with cosmic history, it
 struggles to explain the dynamic character of DE and late-time acceleration without fine-tuning. Prominant examples of MOG
comprises the CDM \cite{nq1} model, $F(R)$ gravity theory \cite{mxxx1,mxxx2,mxxx3}, $f (T)$ gravity theory \cite{mxxx4,mxxx5}, $f (R, T)$ gravity theory \cite{nq2}, Rastall gravity \cite{mxxx6,mxxx7} and $f (Q)$
 gravity theory \cite{nq3}. The effects of galactic rotation curves \cite{nq7} dark components
 \cite{nq6}, and the accelerated expansion of the Universe  \cite{nq8}, while in the cases of the $\Lambda$CDM framework \cite{nq9,nq10} interprets DE via cosmological constant within GR, aligning well with the present observations. Although this standard cosmological framework effectively accounts for nearly all cosmic evolution, it still faces the prominent cosmological constant issue, which concerns the extremely small value of the observed constant in contrast to the Universe’s present critical density.

As a simple extension of GR, $F(R)$-gravity substitutes the Ricci scalar $R$ with a generic function $R$, and is proposed to account for cosmic acceleration and structure formation without depending on dark components such as DE or dark matter  \cite{mx3,mx4,mx5,mx6}. The $F(R)$ gravity model aligns with both post-Newtonian and Newtonian limits \cite{mx7,mx8}, and its action incorporates some key aspects of higher-order gravity \cite{mx9}. Through some consistent formulations \cite{NojiriO2003,NojiriO2011}, the modified $F(R)$ gravity framework offers a viable description of the universe's evolution and provides an explanation for numerous cosmological and astrophysical observations  \cite{Mod1,Mod3,mx4,Mod6,Mod7,Mod7a,Mod7b}. Conversely, nonlinear electrodynamics (NLED) extends classical Maxwell theory by incorporating nonlinear corrections to describe electromagnetic phenomena in extreme field regimes, such as those near compact astrophysical objects or in quantum vacuum effects. Pioneered by Born and Infeld in 1934 \cite{mxx1}, NLED was initially developed to resolve the divergence of point charge self-energy in classical electrodynamics, introducing a finite field strength limit.

Modified Maxwell (ModMax) theory introduced in Ref.~\cite{mx10} extends the Maxwell electrodynamics nonlinearly while preserving conformal invariance. This theory is characterized by a dimensionless parameter $(\gamma)$, which smoothly interpolates between Maxwell’s linear theory ($\gamma=0$) and a nonlinear regime, while maintaining invariance under conformal transformations and electric-magnetic duality rotations \cite{mx10}. Unlike other nonlinear extensions, ModMax is uniquely determined via these symmetries, and offers a consistent method for investigating the conformal invariance beyond the realm of classical electrodynamics \cite{mx10}. A lot of work has been done to examine numerous features of ModMax electrodynamics, and the associated solutions of black holes (BHs) as described in Refs.~\cite{md8,md9,md11,md12,md13,md14,md15}.

Further, ModMax theory has garnered attention for its applications in high-energy physics, cosmology, and holography, particularly as a candidate for describing photon interactions in strong electromagnetic fields \cite{mxx5,mxx6,mxx7,mxx8}. This innovative theory has recently received significant attention, encompassing diverse applications such as the thermodynamic structure of magnetic BHs \cite{mxx10}, $F(R)$-Euler–Heisenberg theory \cite{mxx12}, shadow and quasinormal modes \cite{mxx13,mxx14}, 4D Lovelock theory coupled to ModMax \cite{mxx15}, $F(R)$ gravity coupled to ModMax \cite{mxx16}, and accelerated BHs in ModMax \cite{md15}. { Recently, ModMax electrodynamics has received growing attention due to its conformal symmetry and applicability to strong-field physics, motivating investigations into BH configurations, thermodynamics, quasinormal modes, and optical characteristics in both standard and extended modified gravity theories \cite{EslamPanah:2024lbk,Siahaan:2024ioa,Barrientos:2024umq,Guzman-Herrera:2024fkg,EslamPanah:2024fls,Shahzad:2024ljt,Russo:2024xnh,Ayon-Beato:2024vph,Guzman-Herrera:2023zsv,Bakhtiarizadeh:2023mhk,Rathi:2023vhw,Flores-Alfonso:2020nnd,BallonBordo:2020jtw,Flores-Alfonso:2020euz,Sekhmani:2025epe,Anand:2025iwc,Sekhmani:2025jbl,EslamPanah:2025bfh,Baptista:2025ogh,Diaz:2025zuc,Barrientos:2025rde,Yasir:2025xgh,Heidari:2025llu,NooriGashti:2025jmi,Barbosa:2025smt,Hale:2025veb,Bokulic:2025usc,EslamPanah:2024gxx,Kurbonov:2026oem,EslamPanah:2025oqy,Alloqulov:2025dqi,Sekhmani:2025gvv,Jafarzade:2024zqq}. These studies reveal the diverse phenomenological features arising from nonlinear electromagnetic modifications and their interactions with gravitational effects.

On the other hand, the quintessence field (QF) represents a dynamical DE framework in which the equation of state parameter $w$ links the field’s pressure $p_q$ and energy density $\rho_q$ through the following expression $p_q =\rho_q\,  w_q$ constrained by $w_q=[-1,-1/3]$. This formulation has been notably highly effective in addressing the phenomenology associated with DE  \cite{isz16,isz17}. When embedded in BH spacetimes, the QF significantly alters their asymptotic structure, frequently giving rise to additional cosmological-scale horizons \cite{isz18,isz19}. The field is parametrized by two key quantities: (i) a normalization factor
$c$, determining the influence due to the existence of QF on spacetime, and (ii) the state parameter $w$, which defines the equation of state \cite{isz20,isz21}.

 The field of BH thermodynamics unites the principles of thermodynamics, quantum mechanics, and general relativity (GR). By interpreting BH entropy in terms of statistical mechanics and associating it with the microscopic configurations around the event horizon, this technique provides an effective way to explore the principles of quantum gravity \cite{1,2}. Their findings revealed a link between the BH’s temperature and the event horizon’s area, as well as between its entropy and the surface gravity. Several theoretical models and analytical approaches have been suggested to investigate these phenomena, leading to significant research across diverse gravitational backgrounds \cite{3,4,5,6,7,8,9,10,11}. One of the most notable developments in this perspective is the Page–Hawking transition \cite{12}, which reveals the deep correlation between thermal radiation and the emergence of BHs in AdS spacetime. This phenomenon is interpreted as the gravitational analog of the confinement–deconfinement phase transition observed in gauge theories within the AdS/CFT correspondence. The study of BH thermodynamics reveals remarkable analogies to classical thermodynamic phenomena, most notably the correspondence between the large–small BH transition and the van der Waals gas–liquid phase transition \cite{13,14,15,16,17}.

 By investigating the conserved charge connected to time-translation symmetry, we can relate the first law of thermodynamics to the total energy (mass) of the BH, using an analogy consistent with classical thermodynamic systems \cite{Al-Badawi:2024mco}. To ensure physical viability, one must examine the BH's response to minor perturbations by studying both its dynamical and thermodynamic stability. Beyond standard thermodynamic techniques, the use of geothermodynamics \cite{Weinhold:1975xej,Ruppeiner:2008kd,Ruppeiner:1995zz} enables the exploration of microscopic interaction structures and provides an independent diagnosis of phase transition through the window of the curvature singularities (for more details regarding the correspondence between the phase transition and different geothermodynamic methods, check Refs.~\cite{NaveenaKumara:2020biu,Quevedo:2008ry,Akbar:2011qw,Hendi:2015cka,Sarkar:2006tg,Bhattacharya:2019qxe,Hendi:2015xya,Banerjee:2016nse,Soroushfar:2020wch}). The shadow study of ModMax BHs solution within $F(R)$-gravity may be utilized to explore thermodynamic stability, while the system’s response to variations in temperature, energy, and other thermodynamic parameters near equilibrium helps assess dynamic stability. The thermodynamic behavior of BHs subjected to perturbations can be examined through micro-canonical, canonical, and grand canonical ensembles, as discussed in recent research focusing on thermal stability and related properties \cite{Bardeen:1973gs,Hawking:1974rv,Strominger:1997eq,Hendi:2020mhg,Al-Badawi:2024mco,Nashed:2025ebr}.

In contrast to ordinary blackbody radiation, Hawking radiation manifests as separated quanta—an effect called radiation sparsity, indicating that, despite a constant Hawking temperature, the emission occurs in discrete rather than continuous spectra \cite{Page:1976ki}. The sparse emission characteristic of Hawking radiation is essential in shaping its detectability and implications for quantum information. It can be confirmed that the sparsity parameter is invariant under modifications in particle spin or spacetime dimension, signaling that the quantized nature of Hawking radiation is a general characteristic \cite{Gray:2015pma}.

Motivated by the challenge of detecting strong-field observational and thermodynamic aspects of the alternative theories of GR, we develop a new exact BH solution in higher-order curvature gravity by coupling quintessence DE to the ModMax nonlinear electrodynamics. Our solution offers a non-trivial generalization of previously known ModMax BHs by simultaneously incorporating higher-order curvature modifications and a dynamical DE component, while smoothly reducing to well-known solutions under appropriate limits.  Although ModMax BHs and higher-order curvature theories have each been studied independently, their combined impact on thermodynamic behavior, microscopic phase transitions, and optical properties remains unexplored. In particular, the combined effects of DE and higher-order curvature corrections on the formation of photon spheres and BH shadows have not yet been systematically explored. By adopting Wald's entropy framework in our thermodynamic and geothermodynamic analysis, we address stability, microscopic interactions, and phase transitions, while the optical features of our BH solution are examined through null geodesic analysis and black hole shadows. In the context of optical properties, we determine an exact analytical solution for the photon sphere radius for different values of the quintessence parameter $(\omega)$, allowing us to clearly observe the effects of DE on the optical properties of our BH solution. Thereby, we observe that the combined effect of higher-order curvature terms and DE yields unique, potentially detectable deviations in BH shadows, highlighting their potential as probes of other higher-order curvature theories in future observational research. Furthermore, the BH shadow is naturally characterized from the viewpoint of a static observer at a finite distance owing to the absence of asymptotic flatness. The arrangement of this paper is given as: Sec.~\ref{sec2} presents the formulations of the Einstein field equations
 by combining ModMax NLED and $F(R)$ gravity in the presence of QF and determines the spherically symmetric BH solution. In Sec.~\ref{sec4}, we will study
 the thermodynamic behavior of the system and verify the first law of thermodynamics. Furthermore, we will also compute the entropy of our BH solution using Wald's entropy framework. Sec.\ref{sec6} focuses on analyzing the sparsity of Hawking radiation and the rate of energy emission governed by the entropy. Next, in  Sec.~\ref{sec3}, we investigate the null geodesics to
determine photon sphere and shadow radii. Further,  analytical
expressions for these radii are obtained and plotted for parameter variations. Finally, we presented our concluding remarks in Sec.~\ref{sec5}.

\section{ModMax BHs in $F(R)$-gravity with Quintessence} \label{sec2}

In this section, we develop a novel exact BH solution within higher-order curvature gravity (which, in our case, is $F(R)$-gravity) by incorporating ModMax nonlinear electrodynamics in the presence of a quintessence DE field. This approach provides a unified gravitational framework that combines higher-order curvature corrections, nonlinear electrodynamics, and DE effects. In contrast to previous studies, which mostly focused on ModMax BH solutions in general relativity or higher-order curvature gravity without considering DE, our BH solution model incorporates the effects of these components in both the near-horizon regions and the asymptotic geometry. Furthermore, our BH solution generalizes well-known BH models reported in Ref.~\cite{mxx16,mxx14,mxx10}, enabling a systematic investigation of the roles of DE and higher-order curvature terms in shaping thermodynamic characteristics, microscopic phase structure, and optical properties. However, in the corresponding limiting cases, our solution naturally provides the well-known BH spacetimes, thereby validating its physical consistency. Now, firstly, we mathematically express  action by considering the ModMax and $F(R)$-gravity theoretical framework in four-dimensional spacetime, with a quintessence background as follows
\begin{equation}
\mathcal{I}=\frac{1}{16\pi }\int_{\partial \mathcal{M}}d^{4}x\sqrt{-g}%
\left[F(R)-4(\mathcal{L}_\mathrm{QF}+\mathcal{L}_\mathrm{MM})\right] ,  \label{actionF(R)}
\end{equation}%
where $g$ denotes the determinant of the metric tensor $(g_{\mu \nu})$, while $F(R)=f\left( R\right)+R$, where $R$ is the Ricci scalar and $f\left( R\right)$ represents a function of it. Furthermore, the symbols $\mathcal{L}_\mathrm{MM}$ and $\mathcal{L}_\mathrm{QF}$ are associated with the Lagrangians of the ModMax field and quintessence component, respectively. As described in Refs.~\cite{mx10,mxx20} one can defined the ModMax Lagrangian $\mathcal{L}_\mathrm{MM}$, as follows
\begin{equation}
\mathcal{L}_\mathrm{MM}=\frac{1}{2}\left(\cosh\!\gamma \ \mathcal{S} -\sinh\!\gamma \ \sqrt{\mathcal{P}^{2}+%
\mathcal{S}^{2}} \right) ,  \label{ModMaxL}
\end{equation}%
where the dimensionless parameter in ModMax theory is presented by $\gamma$ and $\mathcal{P}$, and $\mathcal{S}$ are,
respectively, a pseudoscalar and a true scalar, which take the following forms
\begin{equation}
\mathcal{S}=\frac{\mathbb{F}}{2},~~~\&~~~\mathcal{P}=\frac{\widetilde{%
\mathbb{F}}}{2},
\end{equation}%
where $\mathbb{F}=F^{\mu \nu}F_{\mu \nu } $ denotes the Maxwell invariant ($F_{\mu
\nu }=-\partial _{\nu }\mathbb{A}_{\mu }+\partial _{\mu }\mathbb{A}_{\nu }$ (Here, $\mathbb{A}_{\mu }$
presents the gauge potential) is the electromagnetic tensor). Moreover, $%
\widetilde{\mathbb{F}}$ equals to $F_{\mu \nu }\widetilde{F}^{\mu \nu }$,
where $\widetilde{F}^{\mu \nu }=\frac{1}{2}\epsilon _{\mu \nu }^{~~~\rho
\lambda }F_{\rho \lambda }$. By setting $\gamma =0$, the ModMax theory loses its nonlinear electromagnetic nature and reverts to the linear form inherent in Maxwell’s theory. \\ Similarly, by using the method given in Ref.~\cite{Ghosh:2017cuq}, we defined the  Lagrangian $\mathcal{L}_\mathrm{QF}$, which is given as
\begin{equation}
    \mathcal{L}_\mathrm{QF}=-\left(\mathcal{V}(\phi)+\frac{1}{2}\left(\nabla\phi\right)^2\right)\,,
\end{equation}
with the quintessence scalar field $\phi$, and the potential assigned to the QF is $\mathcal{V} (\phi)$. Although we introduce the scalar-field Lagrangian for completeness, our construction follows the effective fluid approach following Ref.~\cite{Ghosh:2016cuq}, where the quintessence is characterized by its equation of state rather than by an explicit scalar potential.

For the $F(R)$ model associated with action (\ref{actionF(R)}), an electrically charged BH can be obtained by focusing explicitly on the source of the electric field within the ModMax theory and plugging $\mathcal{P}=0$. Notably, when considering purely electric configuration ($\mathcal{P}=0$), the ModMax Lagrangian becomes significantly simpler because it depends on the electromagnetic invariants $\mathcal{S}$ and $\mathcal{P}$ \cite{mx10}. In this limit, removing the pseudoscalar invariant leads to a reduction in the theory's nonlinear feature. This behavior is consistent with the established nature of nonlinear electrodynamics, in which confined field configurations can effectively limit nonlinear effects. Therefore, although the parameter $\gamma$ continues to impact the electromagnetic sector, a genuinely nonlinear description of ModMax electrodynamics requires the inclusion of both invariants, as highlighted in duality-invariant frameworks \cite{md14,md9}. In addition, our approach is consistent with earlier investigations of ModMax nonlinear electrodynamics \cite{mx10,mxx16,EslamPanah:2024lbk,EslamPanah:2024fls,Jafarzade:2024zqq}, in which the theory is expressed in terms of electromagnetic invariants and reduces to a simpler form in particular field configurations.   Thereby, the action given in Eq.~\eqref{actionF(R)} yields the following field equations
\begin{eqnarray}
\left( 1+f_{R}\right) \ R_{\mu \nu } -\frac{F(R)\ g_{\mu \nu }}{2}+ f_{R}\left( g_{\mu
\nu }\nabla ^{2}-\nabla _{\nu }\nabla _{\mu }\right)  &=&8\pi \left(T_{\mu\nu}^{\mathrm{MM} }-T_{\mu\nu}^\mathrm{QF}\right),  \label{EqF(R)1} \\
&&  \notag \\
\partial _{\mu }\left( \sqrt{-g}\widetilde{E}^{\mu \nu }\right) &=&0,
\label{EqF(R)2}
\end{eqnarray}%
where $f_{R}=\frac{df(R)}{dR}$ and Ricci tensor denoted by $\mathcal{R}_{\mu\nu}$. Equation~\eqref{EqF(R)1} expresses the gravitational field equations for higher order curvature gravity coupling with matter field, derived through variation of the action with respect to the metric tensor. The leading term on the left-hand side represents a generalized Einstein tensor that captures the higher-order curvature effects by the function $F(R)$, while the terms containing $F_{R}$ describe the dynamical impact of curvature corrections beyond GR. The presence of higher-order derivatives in $F(R)$-theory gives rise to these additional terms, which significantly modify the effective gravitational behavior even for the vacuum solution. The energy-momentum tensor on the right-hand side incorporates contributions from ModMax nonlinear electrodynamics and the quintessence DE component. The presence of nonlinear electrodynamics modifies the traditional Maxwell stress-energy tensor via electromagnetic self-interaction corrections, while the quintessence term provides an effective description of a dynamical DE fluid with state parameter $(\omega)$. Thereby, Eq.~\eqref{EqF(R)1} captures the intricate interplay of nonlinear electromagnetic field, higher-order curvature contributions, and DE, serving as the foundational framework for the BH solution in this work. Additionally, as discussed in Ref.~\cite{isz18}, the energy-momentum tensor corresponding to the ModMax theory, and QF is represented by $T_{\mu\nu}^{\text{MM}}$ and $T_{\mu\nu}^{\text{QF}}$, respectively, and these tensors can be written as
\begin{equation}
4\pi\, T_{MM}^{\mu\nu}=\left( F^{\mu \sigma }F_{~~\sigma }^{\nu
}e^{-\gamma }\right) -e^{-\gamma }\mathcal{S}g^{\mu \nu },  \label{eq3}
\end{equation}
\begin{equation}
    4\pi \,T^{\mu\nu}_{QF}=\nabla_\mu\phi\nabla_\nu\phi-\frac{1}{2}g_{\mu\nu}\left[(\nabla\phi)^2+2\mathcal{V}(\phi)\right],
\end{equation}
and Eq.~(\ref{EqF(R)2}) defines $\widetilde{E}_{\mu \nu }$ as follows
\begin{equation}
\widetilde{E}_{\mu \nu }=\frac{\partial \mathcal{L}}{\partial F^{\mu \nu }}%
=2\left( \mathcal{L}_{\mathbb{S}}F_{\mu \nu }\right) ,  \label{eq3b}
\end{equation}%
where $\mathcal{L}_{\mathcal{S}}=\frac{\partial \mathcal{L}}{\partial
\mathcal{S}}$. Thus, for the case of the electric charge, the ModMax field equation Eq.~(\ref{EqF(R)2}) takes the given shape
\begin{equation}
\partial _{\mu }\left( F^{\mu \nu } e^{-\gamma }\sqrt{-g}\right) =0.
\label{Maxwell Equation}
\end{equation}
Our goal is to develop a BH solution corresponding to model (\ref{actionF(R)}) by assuming a static, spherically symmetric spacetime as
\begin{equation}
ds^{2}=-g(r)dt^{2}+\left(g(r)\right)^{-1}dr^{2}+\left( d\theta ^{2}+\sin
^{2}\theta d\varphi ^{2}\right)r^{2} ,  \label{Metric}
\end{equation}%
in which the metric function is expressed as $g(r)$. Assuming that the scalar curvature is a constant value $R=R_{0}$ the trace of Eq.~(\ref{EqF(R)1}) then becomes
\begin{equation}
\left(f_{R_{0}+1}\right)R_{0} -2\left(f(R_{0}) +R_{0}\right) =0,
\label{R00}
\end{equation}%
where $f_{R_{0}}=~f_{R_{\left\vert _{R=R_{0}}\right.}}$. Treating the scalar curvature as a constant simplifies the gravitational field equations without compromising the key features of higher-order curvature gravity ($F(R)$-gravity theory). Such constant-curvature configurations are a generic feature of effective gravity theories and provide a tractable framework for probing deviations from GR induced by higher-order curvature corrections. Therefore, we rewriting Eq.~(\ref{R00}) in the form of $R_{0}$ yields
\begin{equation}
R_{0}=\frac{2f(R_{0})}{f_{R_{0}}-1}.  \label{R0}
\end{equation}

Plugging Eq. (\ref{R0}) into Eq. (\ref{EqF(R)1}) yields the equations of motion corresponding to the $F(R)$-ModMax framework with QF, which can be expressed in the given manner
\begin{equation}
R_{\mu \nu }\left( 1+f_{R_{0}}\right) -\frac{g_{\mu \nu }}{4}\left(
1+f_{R_{0}}\right)R_{0} =8\pi \left(T_{\mu\nu}^{\text{MM} }-T_{\mu\nu}^{QF}\right).  \label{F(R)Trace}
\end{equation}

The gauge potential is selected in the following form to yield a radial electric field
\begin{equation}
\mathbb{A}_{\mu }=h\left( r\right) \delta _{\mu }^{t},
\end{equation}%
By employing the above-mentioned gauge potential along with Eqs.~(\ref{Maxwell
Equation}) and (\ref{Metric}), yields
\begin{equation}
h(r)=-q/r,  \label{h(r)}
\end{equation}
where $q$ represents an integration constant which corresponds to the electric charge.

The solution of the field equations as described in Ref.~\cite{mxx16}, gives the metric function $g(r)$ in the following form
\begin{equation}
g(r)=1-\frac{m_{0}}{r}-\frac{R_{0}r^{2}}{12}+\frac{q^{2}e^{-\gamma }}{\left(
1+f_{R_{0}}\right) r^{2}}-\frac{c}{r^{3w+1} },  \label{g(r)F(R)}
\end{equation}%
where $m_{0}$ represents an integration constant and $c$ is the quintessence matter normalization factor. The form of metric function given in Eq.~\eqref{g(r)F(R)} reflects the combined influence of mass, nonlinear electromagnetic effects, higher-order curvature modifications (which in our case is related to $F(R)$-gravity), and the surroundings of DE. We should restrict ourselves to the case $%
f_{R_{0}}\neq -1$ in order to get the physical solutions. The term proportional to $m_{0}/r$ describes the usual mass contribution to the gravitational field, whereas the charge-dependent term incorporates the ModMax nonlinear effect that can be reduced to the Maxwell result when we put $\gamma=0$. The curvature corrections dictated by $f_{R_{0}}$ modify the effective electromagnetic coupling and induce geometric deformation of spacetime in the absence of matter fields. The presence of quintessence yields a radial modification governed by the state parameter $(\omega)$, which affects the asymptotic regime and may produce cosmological horizons. While similar solutions have been explored in previous studies \cite{mxx14,mxx16,mxx10}, our approach provides a significant extension by integrating higher-order curvature terms, ModMax electrodynamics, and quintessence dark energy into a single generalized framework. For example, the generalized spacetime described in Eq. (\ref{g(r)F(R)}) encompasses numerous well-known BH solutions as special cases:
\begin{eqnarray*}
   &c=0 \Rightarrow g(r)= 1-\frac{m_{0}}{r}-\frac{R_{0}r^{2}}{12}+\frac{q^{2}e^{-\gamma }}{\left(
1+f_{R_{0}}\right) r^{2}} \Rightarrow
 \text{ $F(R)$-ModMax} \cite{mxx16},
\\& f_{R_0}=0 \Rightarrow g(r)= 1-
 \frac{m_{0}}{r}-\frac{R_{0}r^{2}}
 {12}+\frac{q^{2}e^{-\gamma }}{
 r^{2}}-\frac{c}{r^{3w+1} }
 \Rightarrow \text{ ModMax with QF}
 \cite{mxx14},\\& f_{R_0}=0=c \Rightarrow g(r)= 1-\frac{m_{0}}{r}-\frac{R_{0}r^{2}}{12}+\frac{q^{2}e^{-\gamma }}{ r^{2}} \Rightarrow \text{ ModMax} \cite{mxx10}, \\ &f_{R_0}=0=c=\gamma \Rightarrow g(r)= 1-\frac{m_{0}}{r}-\frac{R_{0}r^{2}}{12}+\frac{q^{2}}{ r^{2}} \Rightarrow \text{Reissner-Nordstr\"{o}%
m-AdS .}
\end{eqnarray*}

\begin{figure}[t]
\centering
\includegraphics[width=3.0in]{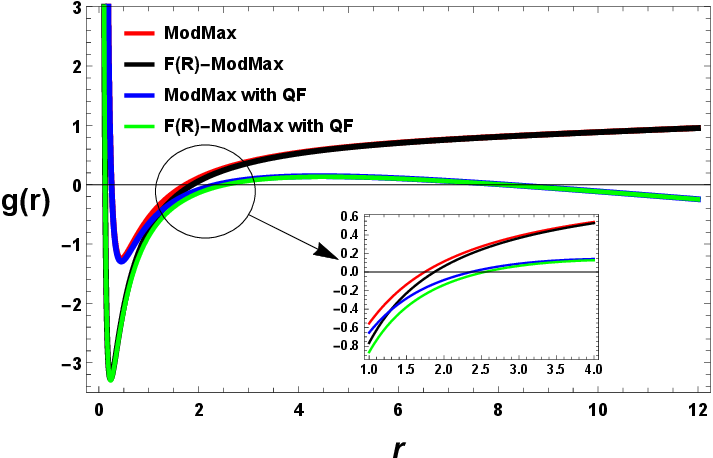}
\includegraphics[width=3.0in]{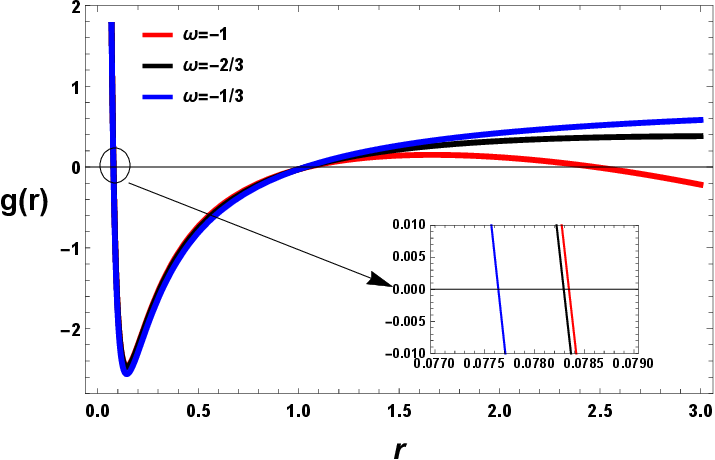}
\caption{\raggedright A comparison of the metric function $g(r)$ for different BH models (left panel) by inserting $m_0=1$, $\gamma=0.6$, $f_{R_0}=0.9$, $q=0.5$, $R_0=-0.01$, $w=-2/3$ and $c=0.1$ for different BH models, such as ModMax (red curve), $F(R)-$Modmax (black curve), ModMax with QF (blue curve) and $F(R)-$ModMax with QF (green curve). To explicitly show the impact of quintessence DE (right panel), we plot the metric function of our BH solution for the state parameters $\omega = -1,~-2/3,~-1/3$, yielding red, black, and blue curves, respectively.}\label{lapseall}
\end{figure}

Fig.~\ref{lapseall} illustrates the comparison of the metric functions corresponding to various BH models (left panel): ModMax (red curve), $F(R)$-ModMax (black curve),  ModMax with QF (blue curve),  and $F(R)$-ModMax with a QF (green curve) by putting $m_{0}=1$, $\gamma=0.6$, $f_{R_0}=0.9$, $q=0.5$, $R_0=-0.01$, $w=-2/3$, and $c=0.1$. One can observe the location of zero points of the metric functions, which directly identify the event and cosmological horizons. The appearance of multiple horizon intersection demonstrates that the spacetime configuration is substantially dictated by the interaction of nonlinear electrodynamic fields, higher-order curvature corrections (owing to the $F(R)$-gravity), and DE. Furthermore, it can also be observed from Fig.~\ref{lapseall} (right panel) that the incorporation of quintessence DE induces notable changes in the global spacetime topology, as the value of the parameter $\omega$ governs the emergence of the cosmological horizon and impacts the causal structure of the BH spacetime. Therefore, Fig.~\ref{lapseall} demonstrates that $F(R)$-gravity, ModMax nonlinear electrodynamics, and quintessence impact the spacetime geometry in qualitatively different ways, providing a clear geometrical framework for the subsequent study of thermodynamics, geothermodynamics, and BH shadows.

Furthermore, we investigate the singularity test of our BH solution by determining the Ricci squared invariant and Kretschmann scalar from the line element given in Eq.~\eqref{Metric}. It is well known that the Schwarzschild BH  possesses a physical singularity at $r=0$ and a coordinate singularity at $r=2M$. We analyze the role of our BH solution within the framework of higher-order curvature gravity ($F(R)$-gravity) in determining the singularity or regular nature of the spacetime at $r=0$. Accordingly, we first determine the Ricci squared and Kretschmann scalar from Eqs.~\eqref{Metric} and \eqref{g(r)F(R)}, which leads to
\begin{eqnarray}\nonumber
    R^{\alpha\beta}R_{\alpha\beta}&=&\frac{1}{4 r^8}\bigg\{18 c \omega  r^{1-6 \omega } \left(c r \omega  \left(9 \omega ^2+6 \omega +5\right)-\frac{4 e^{-\gamma } q^2 (\omega +1) r^{3 \omega }}{f_{R_0}+1}\right)+6 c R_0 \omega  (3 \omega -1) r^{5-3 \omega }+r^8 R_0^2\\\label{RS2s}&+&\frac{16 e^{-2 \gamma } q^4}{\left(f_{R_0}+1\right)^2}\bigg\}\,,\\\nonumber
   K= R^{\alpha\beta\mu\nu}R_{\alpha\beta\mu\nu}&=&\frac{e^{-2 \gamma } r^{-6 \omega -8}}{6 \left(f_{R_0}+1\right)^2} \Bigg[e^{2 \gamma } f_{R_0}^2 r^2 \bigg(18 \left(c^2 U_0+4 c m_{0} \left(3 \omega ^2+5 \omega +2\right) r^{3 \omega }+4 m_{0}^{2} r^{6 \omega }\right)+6 c R_0 (3 \omega -1) \omega  r^{3 \omega +3}\\\nonumber&+&R_0^2 r^{6 \omega +6}\bigg)+18 c^2 e^{2 \gamma } r^2 U_0+2 e^{\gamma } f_{R_0} r \left(6 c e^{\gamma } R_0 (3 \omega -1) \omega  r^{3 \omega +4}+e^{\gamma } R_0^2 r^{6 \omega +7}+18 U_1\right)+72 c e^{\gamma }\\\nonumber&\times& (\omega +1) r^{3 \omega +1} \left(e^{\gamma } m_{0} r (3 \omega +2)-q^2 (9 \omega +4)\right)+6 c e^{2 \gamma } R_0 (3 \omega -1) \omega  r^{3 \omega +5}+24 r^{6 \omega } \bigl(3 e^{2 \gamma } m_{0}^{2} r^2\\\label{KSe}&-&12 e^{\gamma } m_{0} q^2 r+14 q^4\bigr)+e^{2 \gamma } R_0^2 r^{6 \omega +8}\Bigg]\,,
\end{eqnarray}
where
$
U_{0}=27 \omega ^4+54 \omega ^3+51 \omega ^2+20 \omega +4\,$ and
 $U_{1}=c^2 e^{\gamma } r U_{0}-2 c (\omega +1) r^{3 \omega } \left(q^2 (9 \omega +4)-2 e^{\gamma } m_{0} r (3 \omega +2)\right)+4 m_{0} r^{6 \omega } \left(e^{\gamma } m_{0} r-2 q^2\right)\,$. It is evident from Eqs.~\eqref{RS2s} and \eqref{KSe} that the behavior of Ricci squared and Kretschmann scalar is divergent at $r=0$, and also we graphically present their behavior in Fig.~\ref{RSKSe} by employing $m_0=1$, $\gamma=0.6$, $f_{R_0}=0.9$, $q=0.5$, $R_0=-0.01$, $w=-2/3$ and $c=0.1$ for different values of the state parameters $\omega = -1$ red (curve),~$\omega =-2/3$ (black curve),~$\omega =-1/3$ (blue curve). This indicates that our BH solution does not impact the presence of physical singularity at $r=0$. Moreover, the above equations clearly demonstrate that the scalar quantities gradually vanish as $r\to \infty$. This indicates that our solution is unique and preserves the singularity.
\begin{figure}[t]
\centering
\includegraphics[width=3.0in]{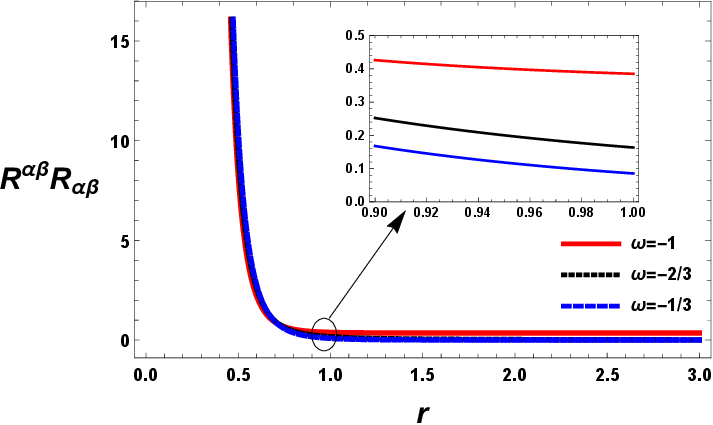}
\includegraphics[width=2.74in]{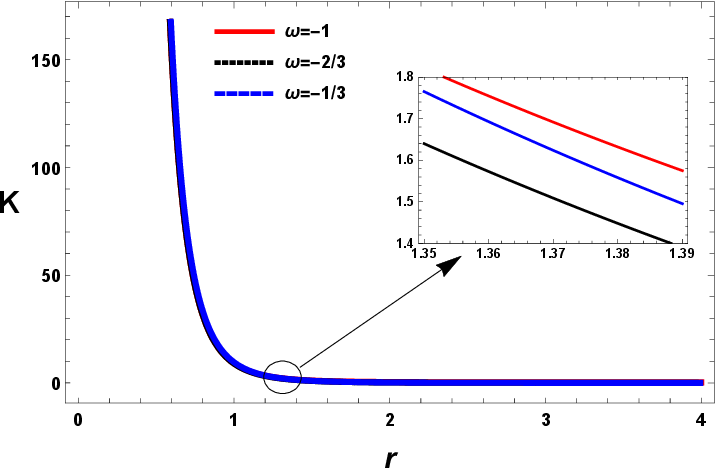}
\caption{\raggedright Behavior of Ricci squared invariant $R^{\alpha\beta}R_{\alpha\beta}$ (left panel) and Kretschmann scalar $K$ (right panel) in term of $r$ is demonstrated. The parameters are set to  $m_0=1$, $\gamma=0.6$, $f_{R_0}=0.9$, $q=0.5$, $R_0=-0.01$, $w=-2/3$, and $c=0.1$. The impact of quintessence state parameters is illustrated for different values $\omega = -1,~-2/3,~-1/3$, with red, black, and blue curves, respectively.}\label{RSKSe}
\end{figure}

\section{Thermodynamics of Black Hole} \label{sec4}
In this section, we study the thermodynamic properties of the new derived ModMax BHs solution within $f(R)$ gravity (higher-order curvature gravity) in the presence of DE. By employing thermodynamic analysis, we can investigate BH stability and phase structure, clarifying how nonlinear electrodynamic fields, curvature corrections, and DE components collectively shape the equilibrium dynamics. First, we compute basic thermodynamic quantities essential to this analysis to probe different thermodynamic aspects of our BH solution and compare it with other BH models. We began our analysis, by using Eq.~\eqref{g(r)F(R)} the expression $g(r_{h})=0$ to compute the mass of the BH as follows

\begin{eqnarray}\label{M1}
M=-\frac{1}{2} \left(f_{R_{0}}+1\right) r_{\rm h} \left(c r_{\rm h}^{-3 \omega -1}-\frac{e^{-\gamma } q^2}{\left(f_{R_{0}}+1\right) r^{2}_{\rm h}}+\frac{r^{2}_{\rm h} R_{0}}{12}-1\right)~.
\end{eqnarray}
where we plugged $m_{0}$ given in Eq.~\eqref{g(r)F(R)} equal to $\frac{2M}{1+f_{R_{0}}}$ which is considered as the total mass of the ModMax BHs as defined in Refs.~\cite{mxx16,Ashtekar:1984zz}. Additionally, the dependence of mass on the ModMax parameter, higher-order curvature terms, and QF highlights how nonlinear electrodynamics contributions and DE dictate the spacetime energy distribution. The curvature-correction parameter plays a key role in rescaling the electromagnetic contribution, where the QF introduces an additional term due to DE.  It is important to mention here that by setting certain parameters, one can obtain the mass function for different ModMax BHs; for example, if we put normalization factor $c=0$, then this mass changes to the $F(R)-$ModMax BH, if we put $f_{R_{0}}=0$ then it becomes the mass of the ModMax with QF field and if we put $c=0$ and $f_{R_{0}}=0$ then it becomes ModMax BH.

 In higher-order curvature gravity theories like our case $F(R)$-theory, the traditional Bekenstein-Hawking entropy relation $S=A/4$ does not strictly hold, as corrections from higher-order curvature terms modify the gravitational action. In this context, the entropy of the BH is determined by Wald's entropy formalism, which relates entropy to the Noether charge associated with diffeomorphism invariance. To obtain the Wald entropy for our BH solution as discussed in Ref.~\cite{Wald:1993nt,Momeni:2025rgk}, we first discuss the general formalism, which can be written as
\begin{eqnarray}\label{WGF}
    S=-2\pi\int_{\rm h}\frac{\delta\mathcal{L}}{\delta R_{\alpha\beta\mu\nu}}\epsilon_{\alpha\beta}\epsilon_{\mu\nu}\sqrt{h}d^{\mathcal{D}-2}x~,
\end{eqnarray}
where $\epsilon_{\alpha\beta}$ corresponds to the binormal vector on the bifurcation surface of the event horizon, defined as an antisymmetric combination of the two null normals to the horizon and  $\sqrt{h}d^{\mathcal{D}-2}x$ defines the infinitesmal surface elements over its cross-section ensuring that the entropy remain purely geometric quantity associated with the horizon. Furthermore, the term $\frac{\delta\mathcal{L}}{\delta R_{\alpha\beta\mu\nu}}$ is the key element within Wald's framework, as it encodes how the gravitational lagrangian varies with the Riemann tensor, thereby providing a mechanism to incorporate higher-order curvature effects, non-minimal coupling, and the alternative theories of gravity. Therefore, this entropy provides a consistent approach to evaluating BH entropy beyond the Bekenstein-Hawking area law, in which the entropy reflects how the gravitational action responds to changes in the spacetime curvature at the horizon. Furthermore, in our case, the relation given in Eq.~\eqref{WGF} simplifies to the given form
\begin{eqnarray}\label{WFFR}
    S=\frac{A}{4}\int_{\rm h}F_{R}(R)dA~.
\end{eqnarray}

By adopting a similar methodology within $F(R)$-gravity as described in Refs.~\cite{mxx16,Nashed:2025ebr,EslamPanah:2024fls}, this naturally produces a modification of the area law, which is given as
\begin{eqnarray}\label{BES}
    S_{\rm HB}=\frac{A}{4}\left(1+f_{R_{0}}\right)~,
\end{eqnarray}
where $f_{R_{0}}$ incorporates the contribution arising from higher-order curvature terms, determined at the constant curvature $R_{0}$. This correction can be interpreted physically as a change in the effective gravitational coupling in $F(R)-$gravity, which consequently rescales the degrees of freedom associated with the horizon. As discussed in Ref.~\cite{Cognola2005}, one can compute the horizon area for the  spherically symmetric spacetime as follows
\begin{eqnarray}\label{ARL}
    A=\int_{0}^{2\pi} \int_{0}^{\pi}\sqrt{g\phi\phi g\theta\theta}=4\pi r^{2}|_{r=r_{\rm h}}=4\pi r^{2}_{\rm h}\,,
\end{eqnarray}
where $r_{\rm h}$ represents the radius of the event horizon. By inserting Eq.~\eqref{ARL} into Eq.~\eqref{BES}, we get
\begin{eqnarray}\label{HB}
    S_{\rm HB}=\pi \left(1+f_{R_{0}}\right) r^{2}_{\rm h}\,.
\end{eqnarray}
This formula offers a physically consistent description of BH entropy in $F(R)-$gravity and recovers the standard area law by taking $f_{R_{0}}\to0$. To express the mass function in terms of the entropy, we start from the entropy expression given in Eq.~\eqref{HB} and, after doing some simple algebra, one can easily derive the following relation in inverted $r_{h}$ form, which is expressed as
\begin{eqnarray}\label{inver}
 r_{h}= \frac{\sqrt{S}}{\sqrt{\pi (f_{R_{0}}+1) }}\,.
\end{eqnarray}

By inserting Eq.~\eqref{inver} into Eq.~\eqref{M1} and simplifying each term, the mass function is explicitly reconstructed as a function of Bekenstein-Hawking entropy (Let us mention here that, by using the same process, one can also reconstruct the thermodynamic mass for different entropy models as described in Refs.~\cite{Jawad:2023ypn,Zafar:2025nho}), as follows
\begin{eqnarray}\label{m2}
M(S,~q)&=&\frac{\sqrt{f_{R_{0}}+1} \sqrt{S}}{2 \sqrt{\pi }} \left\{-c \left(\frac{\sqrt{S}}{\sqrt{\pi  f_{R_{0}}+\pi }}\right)^{-3 \omega -1}-\frac{R_{0} S}{12 \left(\pi  f_{R_{0}}+\pi \right)}+\frac{\pi  e^{-\gamma } q^2}{S}+1\right\}\,.
\end{eqnarray}}
Furthermore, one can obtain the temperature with respect to the entropy by employing Eq.~\eqref{m2} into the following relation
\begin{eqnarray}\label{T1}
    T=\frac{\partial M(S,~q)}{\partial S}&=&\frac{\sqrt{f_{R_{0}}+1}}{16 \sqrt{\pi } S^{3/2}} \left[4 \left(3 c S \pi ^{\frac{3 \omega }{2}+\frac{1}{2}} \omega  \left(\frac{\sqrt{S}}{\sqrt{f_{R_{0}}+1}}\right)^{-3 \omega -1}-\pi  e^{-\gamma } q^2+S\right)-\frac{R_{0} S^2}{\pi  f_{R_{0}}+\pi }\right]\,.
\end{eqnarray}
Here, we note that this relation follows directly from the first law of BH thermodynamics and is obtained by modifying the Bekenstein-Hawking entropy to incorporate higher-order curvature effects. Therefore, one can also verify this fact that this temperature is also equal to the Hawking temperature, which is obtained through the following relation $T=\frac{g'}{4\pi}$ (for more details, regarding the derivation of Hawking temperature in higher-order curvature gravity, check Refs.~\cite {mxx16,Nashed:2025ebr,EslamPanah:2024fls}). In this setting, $M$ represents the system's internal energy due to the absence of the cosmological constant as discussed in Ref.~\cite{Kastor:2009wy}.

According to the first law in BH thermodynamics, changes in the BH’s mass are associated with its fundamental characteristics, including charge, horizon's surface area, and angular momentum. This principle plays a crucial role in examining BH behavior and has remained central in studies, especially within modified gravity frameworks and BH dynamics \cite{Rani:2025esb}.

\begin{eqnarray}\label{Fl}
    dM=TdS+\phi dQ+Adc+BdR_{o}\,,
\end{eqnarray}
where $S,~Q,~R_{0}$ and $A$ are the thermodynamic variables while $T,~\phi,~B$ and $c$ are the conjugate variables associated to these thermodynamic variables. One can also notice the missing terms of thermodynamic pressure and volume, and the reason for this absence, which we mentioned earlier. Therefore, we adopt the same procedure as mentioned in Refs.~\cite{Zafar:2025nho} to obtain the thermodynamic pressure and volume by utilizing entropy, given as
\begin{eqnarray}\nonumber
    V&=&\frac{4 S^{3/2}}{3 \sqrt{\pi } \sqrt{f_{R_{0}}+1}}~,\\\label{PV} P=T\left(\frac{\partial S}{\partial V}\right)&=&\frac{\left(f_{R_{0}}+1\right)}{32 S^2} \left[4 \left(3 c S \pi ^{\frac{3 \omega }{2}+\frac{1}{2}} \omega  \left(\frac{\sqrt{S}}{\sqrt{f_{R_{0}}+1}}\right)^{-3 \omega -1}-\pi  e^{-\gamma } q^2+S\right)-\frac{R_{0} S^2}{\pi  f_{R_{0}}+\pi }\right]~.
\end{eqnarray}
\begin{figure}[t]
\centering
\includegraphics[width=3.0in]{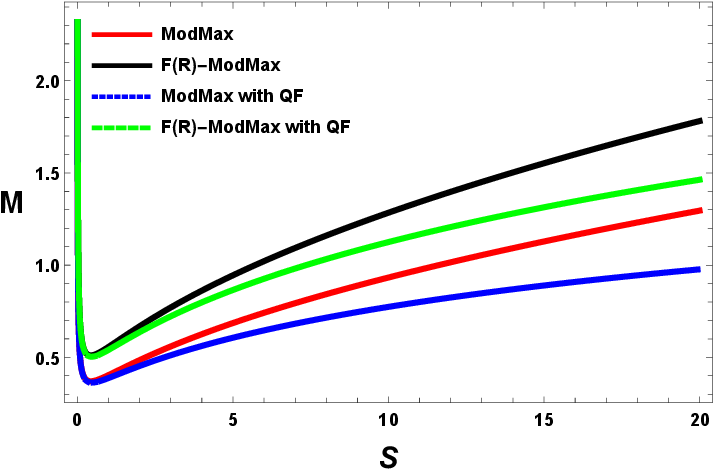}
\includegraphics[width=3.0in]{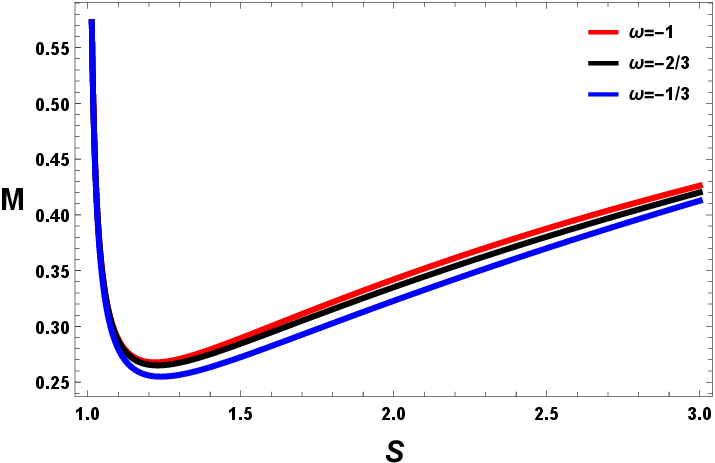}
\caption{\raggedright Plot of mass in terms of entropy $S$ is presented by setting $q=0.5$, $f_{R_0}=0.9$, $c=0.1$, $\gamma=0.6,~R_{0}=-0.01$ and $w=-2/3$ for different ModMax BH models (left panel) such as ModMax BH (red curve), $F(R)-$Modmax BH (black curve), ModMax BH with QF (blue curve) and $F(R)-$ModMax BH with QF (green curve, which is our newly derived BH solution). We demonstrate the mass of our newly derived BH solution (right panel) in the form of entropy for numerous values of $\omega=-1$ (red curve), $\omega=-2/3$ (black curve), and $\omega=-1/3$ (blue curve).}\label{fig-1a}
\end{figure}

\begin{figure}[t]
\centering
\includegraphics[width=3.0in]{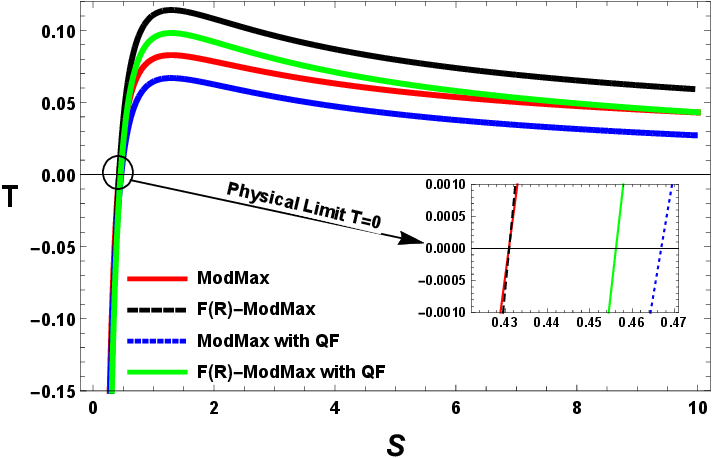}
\includegraphics[width=3.0in]{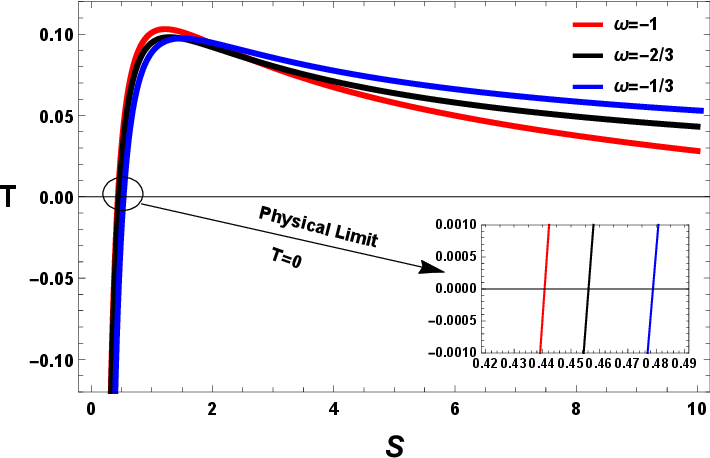}
\caption{\raggedright Plot of $T$ in terms of $S$ is depicted by inserting $\gamma=0.6$, $c=0.1$, $q=0.5$, $f_{R_0}=0.9$, $R_{0}=-0.01$ and $w=-2/3$ for different ModMax BH models such as ModMax BH (red curve), $F(R)$-Modmax BH (black curve), ModMax BH with QF (blue curve) and our solution $F(R)$-ModMax BH with QF (blue curve). Additionally, we presented temperature behavior to observe the influence of QF on our newly derived ModMax BH solution (right panel) in the form of entropy for different values of $\omega=-1$ (red curve), $\omega=-2/3$ (black curve), and $\omega=-1/3$ (blue curve).}\label{fig-1b}
\end{figure}

\begin{figure}[t]
\centering
\includegraphics[width=3.0in]{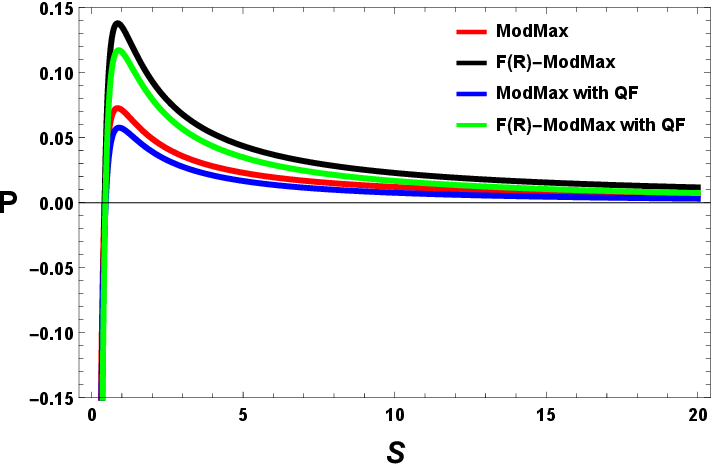}
\includegraphics[width=3.0in]{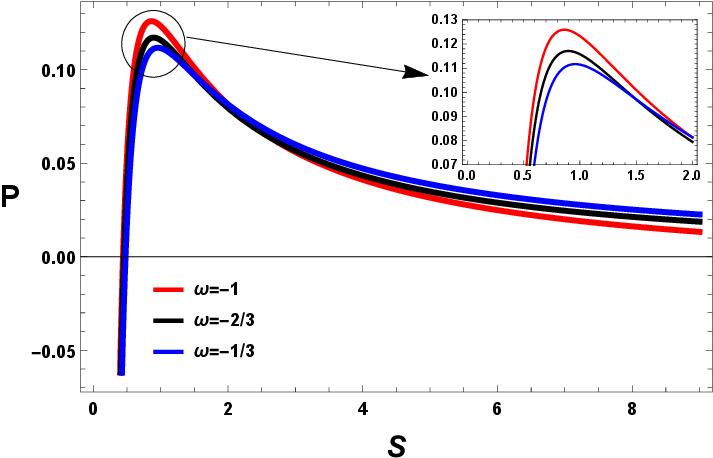}
\caption{\raggedright Plot of pressure in terms of $S$ is depicted by utilizing $f_{R_0}=0.9$,  $q=0.5$, $c=0.1$, $\gamma=0.6,~R_{0}=-0.01$, and $w=-2/3$ for different ModMax BH models such as ModMax BH (red curve), $F(R)-$ModMax BH (black curve), ModMax BH with QF (blue curve), and our $F(R)-$ModMax BH with QF (blue curve). We demonstrated the behavior of the pressure of our newly derived ModMax BH solution (right panel) in the form of entropy for different values of $\omega=-1$ (red curve), $\omega=-2/3$ (black curve), and $\omega=-1/3$ (blue curve).}\label{fig-1c}
\end{figure}

In Figs.~\ref{fig-1a}-\ref{fig-1c}, we present the thermodynamic variables such as mass $m_{0}$, temperature $T$ and thermodynamic pressure $P$ in the form of entropy $S$ by employing $f_{R_0}=0.9$, $q=0.5$, $c=0.1,~\gamma=0.6,~R_{0}=-0.01$ and $w=-2/3$. The trajectories in Fig.~\ref{fig-1a} (left panel) depict different ModMax BH models; for example, ModMax BH (red curve), $F(R)-$ ModMax BH (black curve), ModMax BH with QF (blue curve), and $F(R)-$ModMax BH with QF (green curve, which is our newly derived solution). In Fig.~\ref{fig-1a} (left panel), it can be noticed that initially the mass of the BH decreases, but after some interval of entropy, it begins to increase, and in the ModMax BH with QF depicted that the initial decrease in their behavior is quite rapid as compared to the ModMax BH without QF.  In Fig.~\ref{fig-1a} (right panel), we demonstrate the impact of QF on our BH solution ($F(R)-$ModMax BH with QF) in terms of entropy $S$ for various values of parameter $\omega=-1$ (red trajectory), $\omega=-2/3$ (black trajectory), and $\omega=-1/3$ (blue trajectory). We observe that for smaller values of the QF parameter ($\omega$), the mass of our BH solution increases rapidly as compared to the large values of $\omega$ across all the ranges of entropy. In Fig.~\ref{fig-1b} (left panel), the behavior of the temperature in terms of entropy $S$ is demonstrated by inserting  $f_{R_0}=0.9$,  $q=0.5$, $c=0.1,~\gamma=0.6,~R_{0}=-0.01$ and $w=-2/3$. We observe that the temperature initially shows a strong dependence on small values of entropy, but as entropy increases, the temperature's behavior also increases. Furthermore, the point at which the temperature transitions from negative to positive reflects a possible phase transition in ModMax BHs. In the right panel of Fig.~\ref{fig-1b}, we examine the effect of QF parameter $\omega$ on our BH solution ($F(R)-$ModMax BH with QF) within the framework of entropy $S$ for different values $\omega=-1$ (red trajectory), $\omega=-2/3$ (black trajectory), and $\omega=-1/3$ (blue trajectory). It indicates that temperature rises rapidly to a peak and then declines as entropy continues to increase. Furthermore, we also highlighted that there is a physical limit at which the temperature becomes zero because if the temperature behavior is negative ($T<0$), it reflects that the BH solution is non-physical, and if it is positive ($T>0$), then it is a physical solution. A similar trend is observed in Fig.~\ref{fig-1c} (left panel), which presents the behavior of thermodynamic pressure $P$ in the form of $S$ for different models of ModMax BH, such as ModMax BH (red curve), $F(R)-$ ModMax BH (black curve), ModMax BH with QF (blue curve), and $F(R)-$ModMax BH with QF (green curve). In the right panel of Fig.~\ref{fig-1c}, we study the influence of QF parameter $\omega$ on our BH solution ($F(R)-$ModMax BH with QF) in the form of $S$  for various values $\omega=-1$ (red trajectory), $\omega=-2/3$ (black trajectory), and $\omega=-1/3$ (blue trajectory). We observe that small values of the QF parameter $\omega=-1/3$ suppress other values of $\omega$ in the behavior of thermodynamic pressure $P$ and reach their peak for small ranges of entropy, and then as the entropy grows, large values of $\omega$ dominate. This clearly signifies the impact of the QF parameter on the behavior of thermodynamic pressure.
\subsection{Heat Capacity}
In BH thermodynamics, the amount of energy needed to change the BH’s temperature is known as heat capacity or thermal capacity, a significant and measurable characteristic \cite{davies1978thermodynamics,rodrigues2022bardeen, ruppeiner2014thermodynamic, Tariq:2025wiy}. One can assess BH stability by the sign: positive values indicate stability, and negative values indicate instability. However, the point at which the heat capacity vanishes remains significant, as it reflects possible phase transitions subsequently predicted by geothermodynamic formalisms. By utilizing Eq.~\eqref{T1}, one can compute the expression for the heat capacity as follows
\begin{eqnarray}\label{HC}
    C(S,~q)&=&\frac{2 S \left\{4 \left(3 c S \pi ^{\frac{3 \omega }{2}+\frac{1}{2}} \omega  \left(\frac{\sqrt{S}}{\sqrt{f_{R_{0}}+1}}\right)^{-3 \omega -1}-\pi  e^{-\gamma } q^2+S\right)-\frac{R_{0} S^2}{\pi  f_{R_{0}}+\pi }\right\}}{-12 c S \pi ^{\frac{3 \omega }{2}+\frac{1}{2}} \omega  (3 \omega +2) \left(\frac{\sqrt{S}}{\sqrt{f_{R_{0}}+1}}\right)^{-3 \omega -1}-\frac{R_{0} S^2}{\pi  f_{R_{0}}+\pi }+12 \pi  e^{-\gamma } q^2-4 S}~.
\end{eqnarray}

\begin{figure}[t]
\centering
\includegraphics[width=3.0in]{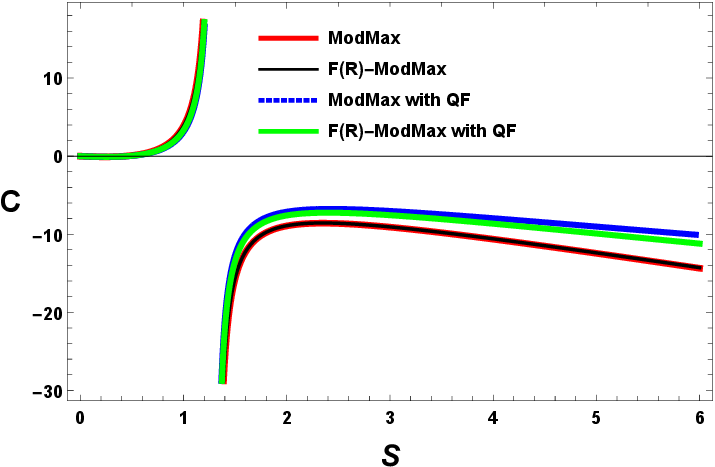}
\includegraphics[width=3.0in]{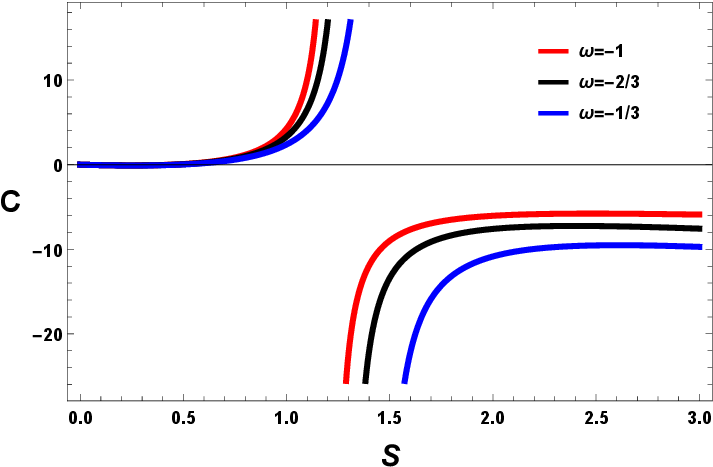}
\caption{\raggedright The behavior of $C$ against $S$ is depicted in Fig.~\ref{fig-2} by inserting $f_{R_0}=0.9$,  $q=0.5$, $c=0.1$, $\gamma=0.6$, $m_{0}=1, R_{0}=-0.01,$ and $w=-2/3$ and we shows various trajectories according to the different ModMax BH models such as ModMax BH (red curve), $F(R)-$Modmax BH (black curve), ModMax BH with QF (blue dotted curve) and our $F(R)-$ModMax BH with QF (green curve). We graphically present the behavior of  $C$ for our ModMax BH solution (right panel) in the form of entropy for different values of $\omega=-1$ (red curve), $\omega=-2/3$ (black curve), and $\omega=-1/3$ (blue curve).}\label{fig-2}
\end{figure}

In Fig.~\ref{fig-2} (left panel), we illustrate the behavior of the heat capacity $C$ in terms of $S$ by plugging $f_{R_0}=0.9$,  $q=0.5$, $c=0.1,~\gamma=0.6,~R_{0}=-0.01$ and $w=-2/3$. The trajectories of $C$ are both negative and positive, suggesting a possible phase transition. It indicates that the trajectories of various ModMax BH are locally unstable for all the values of the horizon radius. Therefore, the behavior of heat capacity reflects a possible phase transition. In the case of the ModMax BH (red curve), there is one zero point of heat capacity placed at $0.4308$, while in the case of the ModMax BH with QF (blue curve), there are three zero points (or divergence points) of heat capacity at $0.4668,~89.286$, and $17502.8$. Similarly, in the context of higher-order curvature gravity ($F(R)$-gravity), the ModMax BH (black curve) has one point of heat capacity at $0.4309$, while in the case of the ModMax BH with QF (green curve), there are three zero points of heat capacity at $0.4562,~170.49$ and $33255.2$. In the right panel of Fig.~\ref{fig-2}, we present the impact of QF on the behavior of $C$ for numerous values $\omega=-1,~-2/3,~-1/3$ for red, black, and blue curves, respectively. We can observe that there are zero points in heat capacity for all the values of QF $\omega$, which highlights the role of this parameter in the local stability of our BH solution in higher-order curvature gravity ($F(R)-$ModMax BH with QF). For example, we find that for boundary values $\omega=-1,-1/3$, the number of zero points of $C$ is two and one, respectively, while for $\omega=-2/3$, the number of zero points of $C$ is three, which we validate through the divergence of the Ricci scalar obtained from the thermodynamic geometry metric formalisms.  Therefore, we found that Fig.~\ref{fig-2} suggests the impacts of $F(R)$ gravity and QF on the thermodynamic stability of the different ModMax BH models.

\subsection{Helmholtz Free Energy}
In thermodynamic systems where temperature and volume remain fixed, the equilibrium behavior of BHs is governed by Helmholtz free energy (HFE). In Refs.\cite{Tariq:2025wiy,paul2024thermodynamics,caneva2021helmholtz,simovic2024euclidean}, Helmholtz free energy is introduced as a means to probe BH's global stability and is mathematically defined as
\begin{eqnarray}\label{HFE1}
    \mathcal{F}&=&M-TS~,
\end{eqnarray}
by employing Eqs.~\eqref{m2}, \eqref{HB}, and \eqref{T1} in Eq.~\ref{HFE1}, we derive the HFE relation for our case, which is given as
\begin{eqnarray}\nonumber
    \mathcal{F}(S)&=&\frac{1}{48 \pi ^{3/2} \sqrt{f_{R_0}+1} \sqrt{S}}\bigg[12 \pi  \left(f_{R_0}+1\right) \left(-3 c S \pi ^{\frac{3 \omega }{2}+\frac{1}{2}} \omega  \left(\frac{\sqrt{S}}{\sqrt{f_{R_0}+1}}\right)^{-3 \omega -1}+\pi  e^{-\gamma } q^2-S\right)-24 \pi  c \left(f_{R_0}+1\right) S \\\label{HFE2}&\times&\left(\frac{\sqrt{S}}{\sqrt{\pi  f_{R_0}+\pi }}\right)^{-3 \omega -1}+24 \pi ^2 e^{-\gamma } \left(f_{R_0}+1\right) q^2+24 \pi  \left(f_{R_0}+1\right) S+R_0 S^2\bigg]~.
\end{eqnarray}

\begin{figure}[t]
\centering
\includegraphics[width=3.0in]{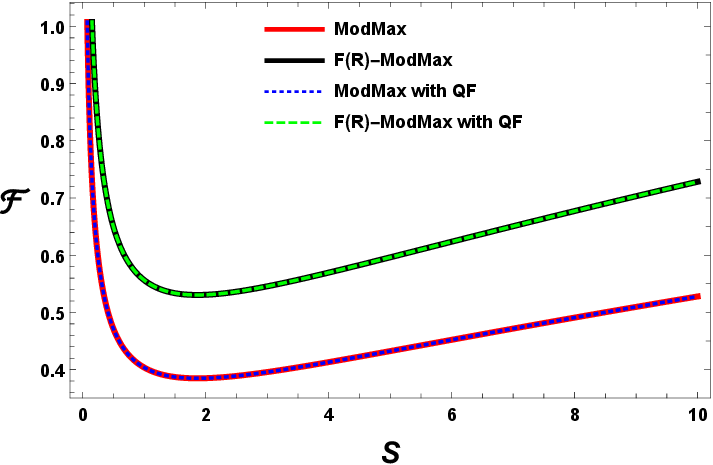}
\includegraphics[width=3.0in]{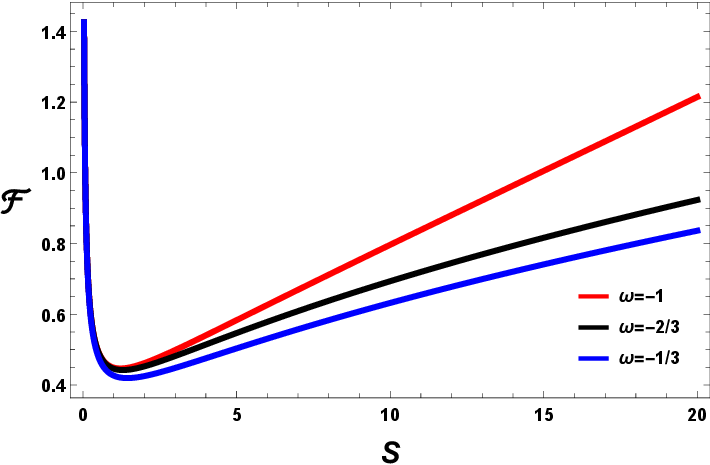}
\caption{\raggedright The behavior of HFE $\mathcal{F}$ as the function of $S$ depicts in Fig.~\ref{fig-3} by inserting $f_{R_0}=0.9$,  $q=0.5$, $c=0.1$, $w=0.6$, $m_{0}=1$, $R_{0}=-0.01,$ and $w=-2/3$. Furthermore, we display various trajectories according to the different ModMax BH models, such as ModMax BH (red curve), $F(R)-$ModMax BH (black dotted curve), ModMax BH with QF (blue curve), and $F(R)-$ModMax BH with QF (green dashed curve). While in the right panel, to examine the impact of QF, we graphically present the behavior of  $C$ for our ModMax BH solution in the form of $S$  for various choices of $\omega=-1$ (red curve), $\omega=-2/3$ (black curve), and $\omega=-1/3$ (blue curve).}\label{fig-3}
\end{figure}

Examining HFE provides valuable insights into the stability of BHs and their potential to undergo phase transitions. In Fig.~\ref{fig-3}, we illustrates the $\mathcal{F}$ in terms of  $S$  by plugging $q=f_{0}=0.9, c=0.1,~\gamma=0.6,~R_{0}=-0.01$ and $w=-2/3$. Fig.~\ref{fig-3} shows that $\mathcal{F}$ is decreasing initially for small values of entropy with positive behavior, and after some interval it begins to increase for large values of entropy. Here, we can also observe that the initial drop of $\mathcal{F}$ is quite sharp for ModMax BH with QF in the absence of $F(R)-$gravity. In our analysis, we observe that the behavior of $\mathcal{F}$ indicates the global stability of all the models of Modmax BHs. Furthermore, in Fig.~\ref{fig-2} (right panel), we present the behavior of $\mathcal{F}$ in terms of entropy $S$ for various values of $\omega=-1,~-2/3,~-1/3$ for red, black, and blue curves, respectively. One can observe that for $\omega=-1$ the behavior of $\mathcal{F}$ initially decreases, but as the entropy increases it becomes to increase sharply as compared to the values $\omega=-2/3$ and $\omega=-1/3$, which highlights the effect of QF in the global stability of our BH solution with the higher-order curvature gravity ($F(R)-$ModMax BH with QF). Moreover, it indicates that, in terms of HFE, smaller values of the QF state parameter $\omega$ dominate over larger ones.

\subsection{Gibbs Free Energy}

We now analyze the Gibbs free energy (GFE) in relation to the entropy. It is well established that the global consistency of BHs can be connected to the GFE, as detailed in \cite{Tariq:2025wiy,ali2019thermodynamics,kubizvnak2012p,deng2018thermodynamics}. The GFE is fundamental in BH thermodynamics, as it reveals key information about their stability and phase transition aspects. The role of GFE is crucial in evaluating the thermodynamic favorability of BHs under constant pressure and temperature conditions. One may compute the GFE for ModMax BH with QF by employing the expression
\begin{eqnarray}\label{GFE}
    \mathcal{G}&=&M-TS+PV~.
\end{eqnarray}

The GFE expressed through entropy $S$ can be examined by substituting Eqs.~\eqref{HB}, \eqref{m2}, \eqref{T1} and \eqref{PV} into Eq.~\eqref{GFE}, which yields
\begin{eqnarray}\nonumber
    \mathcal{G}&=&\frac{}{48 \pi ^{3/2} \sqrt{f_{R_0}+1} \sqrt{S}}\biggl[4 \pi  \left(f_{R_0}+1\right) \left(-3 c S \pi ^{\frac{3 \omega }{2}+\frac{1}{2}} \omega  \left(\frac{\sqrt{S}}{\sqrt{f_{R_0}+1}}\right){}^{-3 \omega -1}+\pi  e^{-\gamma } q^2-S\right)-24 \pi  c \left(f_{R_0}+1\right)S\\\label{GFE1}&\times&  \left(\frac{\sqrt{S}}{\sqrt{\pi  f_{R_0}+\pi }}\right)^{-3 \omega -1}+24 \pi ^2 e^{-\gamma } \left(f_{R_0}+1\right) q^2+24 \pi  \left(f_{R_0}+1\right) S-R_0 S^2\biggr]~.
\end{eqnarray}

\begin{figure}[t]
\centering
\includegraphics[width=3.0in]{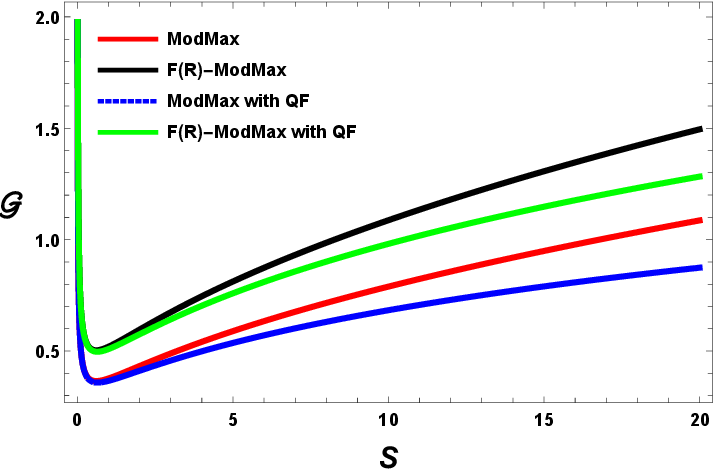}
\includegraphics[width=3.0in]{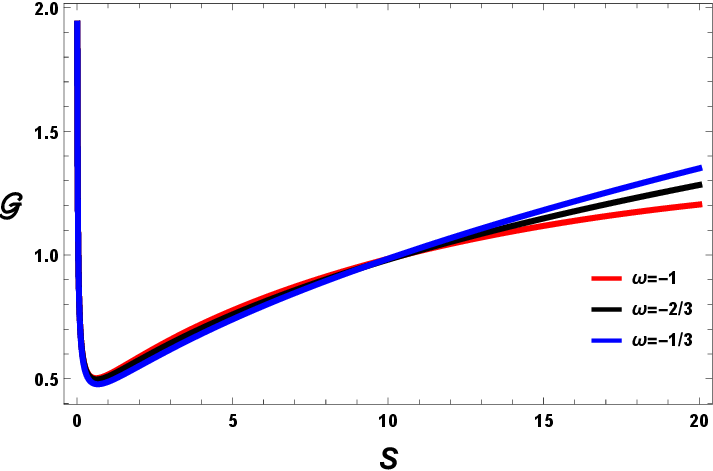}
\caption{\raggedright The behavior of $\mathcal{G}$ in the form of the $S$ demonstrates in Fig.~\ref{fig-4} by inserting $f_{R_0}=0.9$,  $q=0.5$, $c=0.1$, $\gamma=0.6$, $m_{0}=1,~R_{0}=-0.01,$ and $w=-2/3$. Furthermore, we show numerous trajectories with respect to the different ModMax BH models, such as ModMax BH (red curve), $F(R)-$ModMax BH (black curve), ModMax BH with QF (blue dotted curve), and $F(R)-$ModMax BH with QF (blue curve). In the right panel, we present the behavior of  $C$ for our ModMax BH solution in terms of  $S$ for different choices of $\omega=-1$ (red curve), $\omega=-2/3$ (black curve), and $\omega=-1/3$ (blue curve).}\label{fig-4}
\end{figure}

 In Fig.~\ref{fig-4}, we illustrates $\mathcal{G}$ in terms of $S$ by plugging $f_{R_0}=0.9$,  $q=0.5$, $c=0.1,~\gamma=0.6,~R_{0}=-0.01$ and $w=-2/3$. We observe a similar behavior of GFE to that previously observed for HFE. The behavior of GFE is positive for all the values of the entropy in every ModMax BH model, but the impact of QF and $F(R)$ can also be noticed in Fig.~\ref{fig-4}. This indicates that all ModMax BH models are globally stable across all entropy ranges. Although local instability occurs, the HFE and GFE indicate that the BH remains globally stable over a wide range of parameters. This underscores the importance of treating local and global stability independently, particularly in higher-order curvature gravity ($F(R)$-gravity), where additional dynamical degrees of freedom govern the thermodynamic framework. Fig.~\ref{fig-2} (right panel) illustrates the behavior of $\mathcal{F}$ in terms of the $S$ for different values of $\omega = -1, -2/3, -1/3$, represented by red, black, and blue curves, respectively. Surprisingly, unlike the HFE, we observe that larger values of the QF state parameter $\omega$ dominate smaller ones, which is quite interesting. This also reflects the effect of QF on the global stability of our BH solution with higher-order curvature gravity ($F(R) $- ModMax BH with QF).

 It is important to highlight that quintessence dark energy influences the thermodynamic phase structure in a subtle yet physically significant manner. It is evident that its effects on global thermodynamic quantities, including HFE and GFE, are relatively mild, in contrast to its stronger influence on local stability, as reflected in the heat capacity. This difference can be explained through the radial behavior of the quintessence contribution, which follows a scaling of $r^{-3\omega-1}$. When $\omega\in[-1,-1/3)]$, which is physically significant, this term decays slowly with distance and primarily impacts the large-scale structure of the spacetime. At the horizon scale, where mass and charge predominantly determine thermodynamic quantities, the impact of quintessence remains suppressed.
\section{Geothermodynamics in ModMax black holes with QF in $F(R)$-gravity}\label{sec6}
In this section, we discuss the geothermodynamics of our BH solution in higher-order curvature gravity using entropy and investigate the correspondence between the zero point of heat capacity and the curvature scalar in the geothermodynamic formalisms, thereby shedding light on the thermodynamic structure of our BH model. From the perspective of thermal fluctuation theory, geothermodynamics employs the invariant curvature scalar of thermodynamic space to effectively reveal microscopic interactions, critical phenomena, and phase transitions in BHs. By using entropy, we examine and confirm the phase transition in the newly derived ModMax BH in higher-order curvature gravity. We start by introducing the fundamental structure (that is, Weinhold metric formalism) of the geothermodynamic formalism, following the same procedure described in Ref.~\cite{Weinhold:1975xej}, which can be mathematically expressed as
\begin{eqnarray}\label{Wein1}
    g_{\alpha\beta}^{\rm Wein}= \partial_{\alpha}\partial_{\beta}M_{S,q},
\end{eqnarray}
and the corresponding line element of this thermodynamic geometry metric for our BH model is given as
\begin{eqnarray}\label{WeinLE}
   ds^{2}_{\rm Wein}= M_{SS}dS^{2}+2M_{Sq}dSdq+M_{qq}dq^{2}.
\end{eqnarray}
The associated metric to this line element can be expressed as
\begin{equation}\label{WeinMt}
g^\mathrm{Wein}=
\begin{pmatrix}
M_{SS} & M_{Sq}  \\
M_{qS} & M_{qq}  \\
\end{pmatrix}\,.
\end{equation}

The Weinhold metric arises from the  Hessian framework of the mass with respect to the thermodynamic variables, thereby encoding the thermodynamic phase space geometrically. However, as we know that the Weinhold metric framework is often unable to capture the critical behavior in the BH configuration, specifically in the presence of higher-order curvature gravity models, it is employed here primarily as an initial diagnostic approach. Another important geothermodynamic metric formalism is the Ruppeiner metric, which is constructed by making some corrections to the Weinhold metric formalism given in Eq.~\eqref{Wein1}. The physical motivation for this correction is to ensure that the fluctuation theory underlying the Ruppeiner metric is consistent with the thermodynamic probability distribution, which varies with temperature. The presence of the inverse temperature term enhances the interpretation of divergence (or singularities) of the curvature scalar as signatures of critical behavior and phase transitions in BH thermodynamics, making it a basic thermodynamic quantity that dictates the framework of fluctuation geometry and encodes microscopic stability. Following Refs.~\cite{Ruppeiner:2008kd,Ruppeiner:1995zz,Akbar:2011qw,Soroushfar:2020wch}, we may define the framework for the Ruppeiner metric, and its explicit form can be written as
\begin{eqnarray}\label{Ruppein1}
    ds^{2}_{\rm Rup}= T^{^{-1}} ds^{2}_{\rm Wein}\,,
\end{eqnarray}
and the metric associated with Eq.~\eqref{Ruppein1} takes the following shape
\begin{equation}\label{RuppeinMt}
g^\mathrm{Rup}=T^{^{-1}}
\begin{pmatrix}
M_{SS} & M_{Sq}  \\
M_{qS} & M_{qq}  \\
\end{pmatrix}\,.
\end{equation}

\begin{figure}
\centering
\includegraphics[width=3.0in]{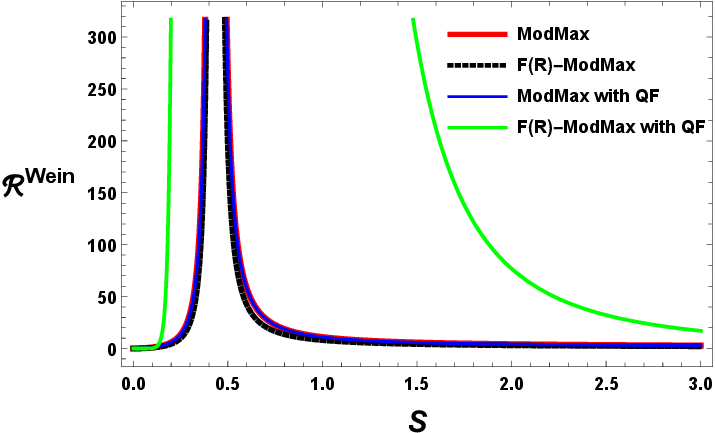}
\includegraphics[width=3.0in]{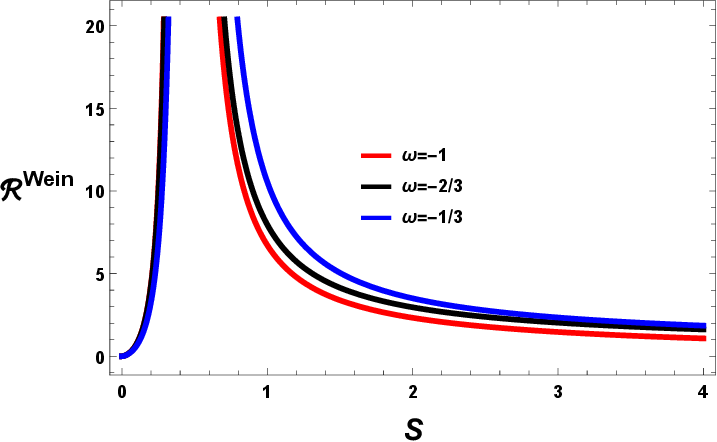}
\includegraphics[width=3.0in]{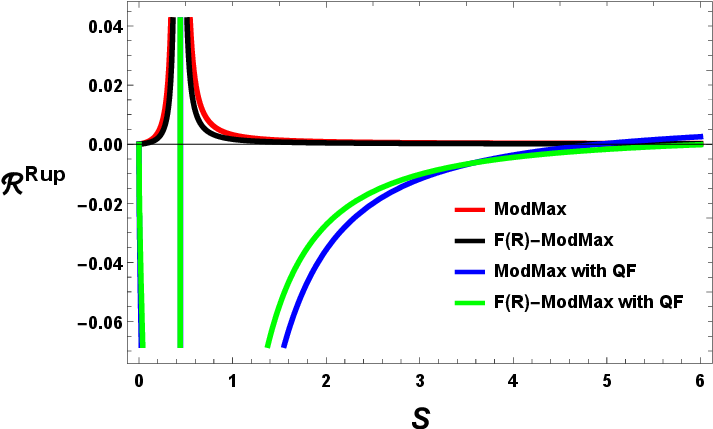}
\includegraphics[width=3.0in]{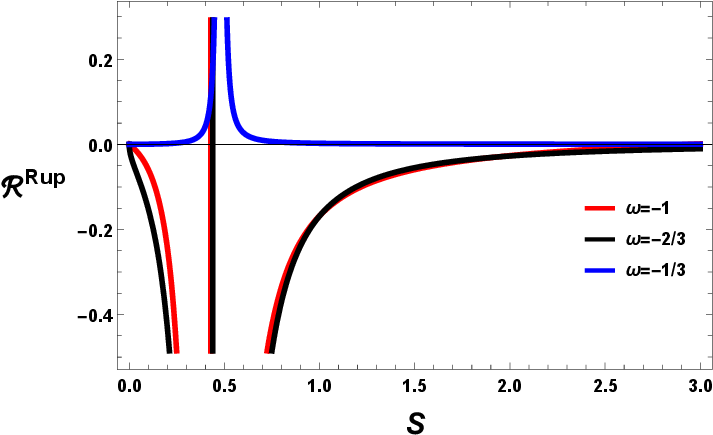}
\caption{\raggedright Plots of $\mathcal{R}^{\rm Wein}$ (upper left panel) and $\mathcal{R}^{\rm Rup}$ (lower left panel) in terms of the $S$ demonstrates in Fig.~\ref{fig-55} by substituting $q=0.5$, $f_{0}=0.9$, $c=0.1,~\gamma=0.6,~R_{0}=-0.01$ and $w=-2/3$. Additionally, for a comprehensive comparison, we obtain different trajectories associated with different ModMax BH models, such as ModMax BH (red curve), $F(R)-$ModMax BH (black curve), ModMax BH with QF (blue curve), and our $F(R)-$ModMax BH with QF (green curve). To examine the impact of DE in our BH solution, we presents the behvaior of $\mathcal{R}^{\rm Wein}$ (upper right panel) and $\mathcal{R}^{\rm Rup}$ (lower right panel) in terms of the $S$ by varying $\omega=-1,~-2/3,~-1/3$ and their trajectories are red, black, and blue, respectively. }\label{WeRU}
\end{figure}

Let us mention that we only present the behavior of the Ricci curvature scalar obtained from the Weinhold and Ruppeiner metrics due to their length. In Fig.~\ref{WeRU}, we presents the behavior of curvature scalars $\mathcal{R}^{\rm Wein},~R^{\rm Rup}$ obtained from the Weinhold and the Ruppeiner metrics in the form of $S$ by putting $q=0.5$, $f_{0}=0.9$, $c=0.1,~\gamma=0.6,~R_{0}=-0.01$ and $w=-2/3$. Moreover, various curves are presented corresponding with various ModMax BH models, such as ModMax BH (red curve), $F(R)-$ ModMax BH (black curve), ModMax BH with QF (blue curve), and $F(R)-$ModMax BH with QF (green curve). From Fig.~\ref{WeRU} (upper left panel), we observe that the behavior of $\mathcal{R}^{\rm Wein}$ is positive for all the values of the entropy, which can be interpreted as the interaction between the particles of all the models of the ModMax BH is repulsive. In addition, we can also notice that there is no coincidence of $\mathcal{R}^{\rm Wein}$ with the zero points of heat capacity for any ModMax BH models, which do not provide any information regarding the phase transition. Furthermore, we demonstrate the behavior of $\mathcal{R}^{\rm Wein}$ for different values of $\omega=-1,~-2/3,~-1/3$ in Fig.~\ref{WeRU} (upper right panel). It can be observed that there is no coincidence of divergence points of $\mathcal{R}^{\rm Wein}$ with the zero points of $C$ for any value of $\omega$.  Therefore, we shift our analysis from the Weinhold metric to the Ruppeiner metric, which is graphically presented in Fig.~\ref{WeRU} (lower left panel). It is observed that $\mathcal{R}^{\rm Rup}$ demonstrates both positive and negative behavior, which indicates the interaction between the particles of the ModMax BH models is both attractive and repulsive. Moreover, we also noticed that in the case of the ModMax BH (red curve), there are five divergence point for $\mathcal{R}^{\rm Rup}$, while in the case of the ModMax BH with QF (blue curve), we obtain five divergence points for $\mathcal{R}^{\rm Rup}$. Similarly, in both cases of higher-order curvature gravity, MadMox BHs with QF (green curve) and ModMax BH without QF (black curve) have five divergence points. Now, here we mention that the divergence of $\mathcal{R}^{\rm Rup}$ in all the cases of the MadMox BH models coincides with all the zero points of heat capacity. In Fig.~\ref{WeRU}, we observed that the divergence of $\mathcal{R}^{\rm Rup}$ is exactly at the same points on which we obtain the zero points of heat capacity for all the values of $\omega$. Furthermore, we also find that for each value of $\omega$, there are multiple divergence points in $\mathcal{R}^{\rm Rup}$ and these divergence points of $\mathcal{R}^{\rm Rup}$ coincide with all the zero points of $C$, which indicates a possible multiple phase transition in our BH solution. Therefore, in our analysis, we observed that the presence of QF and higher-order curvature gravity significantly affects the thermodynamic stability of the ModMax BHs, as evident in Fig.~\ref{WeRU}.

 \section{Sparsity and Energy emission of ModMax black holes with QF in $F(R)$-gravity}\label{sec6}
The emission of Hawking radiation serves as a key probe of the quantum properties of BHs. This section explores the sparsity and energy emission rate for our ModMax BH solution in higher-order curvature gravity with QF (ModMax black holes with QF in $F(R)$-gravity), via entropy $S$. We first assume the BH sparsity framework, noting that their radiation behaves like that of a black body, with the surface gravity dictating the temperature. Hawking radiation departs markedly from standard blackbody radiation, exhibiting an unusually sparse distribution during BH evaporation. The idea of sparsity classifies the average separation between consecutive quantum emissions, which can be computed by their associated energy values as given in Refs.\cite{Page:1976ki,Gray:2015pma}, and its expression takes the following form
\begin{eqnarray}\label{SH}
 \eta =\frac{\mathcal{C}}{\mathfrak{g} }\left(\frac{\lambda_t^2}{\mathcal{A}_\mathrm{eff}}\right),
\end{eqnarray}
where the thermal wavelength is characterized by $\lambda_{t}=2 T^{-1} \pi$, the emitted particle's spin degeneracy is given by $\mathfrak{g}$, the BH's effective area is denoted by $\mathcal{A}_\mathrm{eff}$ and $\mathcal{C}$ is a dimensionless parameter. The massless spin-1 boson emitting in the case of the Schwarzschild BH is $\lambda_t=8\pi r^{2}_{h}$, and its efficiency is  $\eta_\mathrm{SH}=8\pi r^{2}_{h}\approx73.49$. Furthermore, $\eta$ represents the comparison between the characteristic particle emission time and the thermal wavelength scale, where higher values correspond to weak and infrequent Hawking radiation. Furthermore, this parameter responds strongly to variations in entropy and temperature, making it an important tool for examining quantum corrections and DE contributions. In our analysis of sparsity and energy emission of Hawking radiation, we employ the temperature in terms of $S$ for ModMax BHs with QF in $F(R)$-gravity.

\begin{figure}[t]
\centering
\includegraphics[width=3.0in]{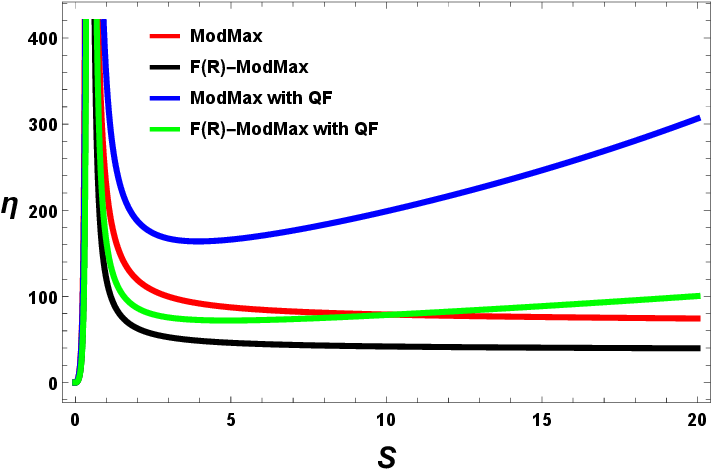}
\includegraphics[width=3.0in]{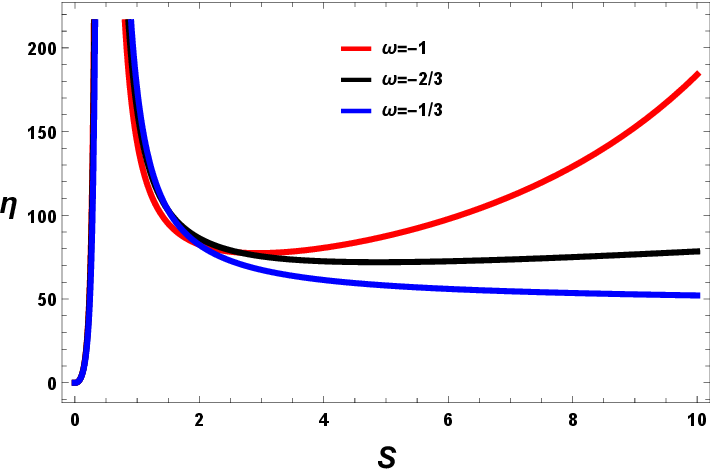}
\caption{\raggedright Plot of $\eta$ in terms of $S$ demonstrates in Fig.~\ref{fig-55} by inserting $q=0.5$, $f_{0}=0.9$, $c=0.1,~\gamma=0.6,~R_{0}=-0.01$ and $w=-2/3$. Moreover, different trajectories are depicted with respect to the different ModMax BH models, such as ModMax BH (red curve), $F(R)-$ ModMax BH (black curve), ModMax BH with QF (blue dotted curve), and $F(R)-$ModMax BH with QF (green dotted curve). In the right panel, we visually illustrate the effect of QF on $\varepsilon_{t\Omega}$ in the form of $S$ for numerous choices of $\omega=-1$ (red curve), $\omega=-2/3$ (black curve), and $\omega=-1/3$ (black curve).}\label{fig-55}
\end{figure}

Fig.~\ref{fig-55} displays how $\eta$ behaves in different ModMax BH models in the form of $S$ by employing $q=0.5,~f_{0}=0.9, c=0.1,~\gamma=0.6,~R_{0}=-0.01$ and $w=-2/3$. For different models of ModMax BH, we obtain different curves, including red, black, blue, and green curves for ModMax, $F(R)-$ModMax BH, ModMax BH with QF, and $F(R)-$ModMax BH with QF, respectively. We notice that QF and $F(R)$-gravity significantly impact the sparsity of the ModMax BHs, which is quite evident in Fig.~\ref{fig-55} (left panel). In context of ModMax BHs with QF in $F(R)$-gravity (green curve) and without $F(R)$-gravity (blue curve), we find that sparsity initially declines for lower values of entropy, but for high ranges of entropy, it begins to increase, and it surpasses the sparsity value in the Schwarzschild case, which reflects that at the evaporation stage the radiation emitted by these BHs is sparser than the Hawking radiation. On the other hand, in the cases of ModMax BHs in the absence of QF with $F(R)$-gravity (black curve) and without $F(R)$-gravity (red curve), one can observe that initially $\eta$ increases for small values of the entropy, but for larger values it begins to drop, and after some large values of entropy the behavior of $\eta$ becomes constant. On the other hand, if we look at Fig.~\ref{fig-55} (right panel), we can clearly observe that for lower values of the parameter $\omega$, $\eta$ increases sharply and it significantly departs from the Schwarzschild case, which indicates that the radiation emitted at the evaporation stage is sparser than the Hawking radiation. Our results show that both higher-order curvature corrections and quintessence significantly influence the sparsity of Hawking radiation, with a larger curvature-correction parameter leading to greater sparsity and suppressing the evaporation process. Similarly, QF increases the interval between successive emission events, thereby suppressing Hawking evaporation and emphasizing the influence of DE and higher-order curvature gravity on BH evaporation.

Hawking radiation leads to the creation of antiparticle-particle pairs near the horizon $r_{h}$, initiating evaporation as positively charged particles escape outwards via quantum tunneling. The rate at which energy is radiated by the distant observer dictates how quickly the BH evaporates. The presence of a large energy-absorption cross section for this observer corresponds to the shadow of the BH, indicating the interaction between the incoming radiation and the BH, while the value of the cross-section in the limiting case agrees with that given in Ref.~\cite{Mushtaq:2025ksk}, and its mathematical expression is given as
\begin{eqnarray}\label{CS1}
\sigma _{\lim }=\pi  r^2_{h},
\end{eqnarray}
Utilizing the above equation, the rate of energy emission can be computed, which is given as
\begin{eqnarray}\label{EQ}
\frac{d^2 \varepsilon }{dtd\Omega }=\frac{2 \Omega ^3 \pi ^2 r_{+} }{e^{\Omega T^{-1}}-1}.
\end{eqnarray}
\begin{figure}[t]
\centering
\includegraphics[width=3.0in]{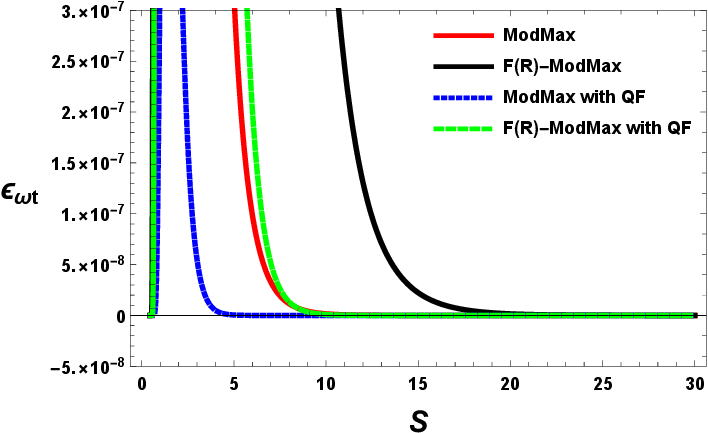}
\includegraphics[width=3.0in]{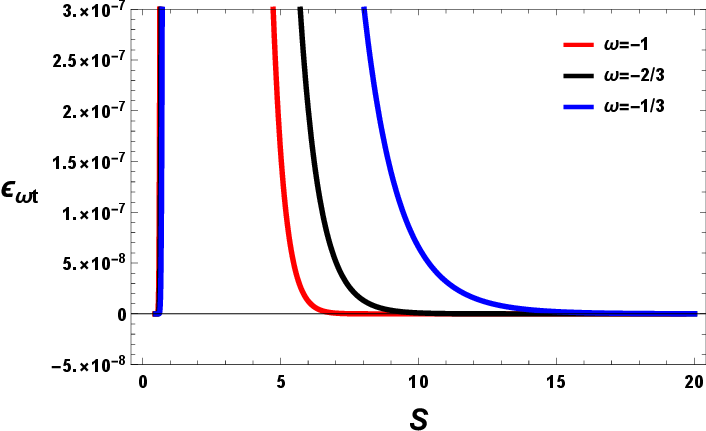}
\caption{\raggedright Plot of $\varepsilon_{t\Omega}$ in the form of the $S$ is illustrated by substituting $q=0.5$, $f_{0}=0.9$, $c=0.1$, $\gamma=0.6,~R_{0}=-0.01,$ and $w=-2/3$. Moreover, different trajectories are shown with respect to the ModMax BH models, such as ModMax BH (red curve), $F(R)-$ModMax BH (black curve), ModMax BH with QF (blue curve), and $F(R)-$ModMax BH with QF (blue curve). In the right panel, we graphically present the impact of QF on $\varepsilon_{t\Omega}$ in the form of $S$ for different values of $\omega=-1$ (red curve), $\omega=-2/3$ (black curve), and $\omega=-1/3$ (black curve).}\label{fig-6}
\end{figure}

Fig.~\ref{fig-6} displays how $\varepsilon_{t\Omega}$ varies in different ModMax BH models in terms of entropy $S$ by plugging $q=0.5,~f_{0}=0.9, c=0.1,~\gamma=0.6,~R_{0}=-0.01$ and $w=-2/3$. For various models of ModMax BH, we get different trajectories, such as red, black, blue, and green curves for ModMax, $F(R)-$ModMax BH, ModMax BH with QF, and  $F(R)-$ModMax BH with QF, respectively. It is observed that in all the models of ModMax BHs, $\varepsilon_{t\Omega}$ is increasing in a very small interval of entropy, which means for larger BHs the energy emission rate is faster. But $\varepsilon_{t\Omega}$ drops for large values of entropy, and eventually it converges to zero in all the cases of ModMax BHs. Although for ModMax BHs without QF in the presence and absence of $F(R)$-gravity, $\varepsilon$ approaches zero for lower values of entropy, in contrast to the ModMax BHs with QF in the presence and absence of $F(R)$-gravity. A similar trend is also observed in Fig.~\ref{fig-6} (right panel), which is obtained by putting $\Omega=-1$ (red curve), $\omega=-2/3$ (black curve), and $\omega=-1/3$ (black curve). By incorporating higher-order curvature corrections, exponential corrections to entropy and QF, our BH model yields a conjugate-temperature evaluation that departs from the classical Hawking spectrum, ultimately reducing the energy emission rate, with the effect more pronounced for smaller BHs.

\section {Shadow of ModMax BH with QF in $F(R)$-gravity} \label{sec3}

The analysis of BH shadows provides a direct window to the strong-field gravitational regime, making them an effective tool for probing deviations from GR. Since the shadow boundary is defined by the unstable circular photon motion, it responds to changes in both the near-horizon geometry and the global spacetime structure. In this section, we investigate the photon sphere and BH shadow of the newly derived ModMax BH in $F(R)$-theory in the presence of QF, seeking to reveal optical signatures corresponding to the nonlinear electrodynamics, curvature corrections, and quintessence effects. Therefore, by employing the same procedure as discussed in Refs.~\cite{AA0,AA1,AA2,AA3,sec3is07}, we compute the geodesic equations by utilizing the function of the Lagrangian density, given as
\begin{equation}
    \mathcal{L}= \frac{1}{2} g_{\mu\nu} \dot{x}^\mu\dot{x}^\nu,\label{bb1}
\end{equation}
in which an overdot denotes differentiation with respect to the affine parameter $\lambda$.

Since the spacetime possess spherical symmetry and by utilizing (\ref{Metric}), the Lagrangian density simplifies when evaluated at ($\theta=\pi/2$ )
\begin{equation}
    \mathcal{L}=\frac{1}{2}\,\left[-g(r)\,\left(\frac{dt}{d\lambda}\right)^2+\frac{1}{g(r)}\,\left(\frac{dr}{d\lambda}\right)^2+r^2\,\left(\frac{d\phi}{d\lambda}\right)^2\right].\label{bb2}
\end{equation}

There are two conserved terms associated with their cyclic coordinates ($t, \phi$), which are expressed as
\begin{equation}
    \mathrm{E}=-\frac{\partial \mathcal{L}}{\partial \dot{t}}=\,\frac{dt}{d\lambda}\,g(r),\label{bb3}
\end{equation}
and
\begin{equation}
    \mathrm{L}=\frac{\partial \mathcal{L}}{\partial \dot{\phi}}=r^2\,\frac{d\phi}{d\lambda},\label{bb4}
\end{equation}
where $\mathrm{L}$ and $\mathrm{E}$ are the conserved angular momentum and energy, respectively. These constants naturally define the impact parameter \cite{bozza_gravitational_2010}
\begin{equation}
    b=\frac{\mathrm{L}}{\mathrm{E}} = \frac{r^2}{g(r)}\frac{d\phi}{d t},
    \label{eq:b_def}
\end{equation}
which plays a central role in determining the boundary of the BH shadow.  For photon trajectories, the condition $\mathcal{L}=0$ governs the dynamics, which leads to the equations
\begin{eqnarray}
    && \dot{r}^{2} + V_{\rm eff}(r) = \mathrm{E}^{2},
    \label{bb5}\\
    && \left( \frac{d\phi}{d\lambda} \right)^{2}
    = \frac{r^{4}}{b^{2}} \left[1 - \frac{V_{\rm eff}(r)}{\mathrm{E}^{2}} \right],
    \label{bb5p}
\end{eqnarray}
where the effective potential governing the motion of incoming photons is given by
\begin{equation}
    V_{\rm eff}(r)
    = \mathrm{L}^{2}\frac{f(r)}{r^{2}}=\frac{\mathrm{L}^{2}}{r^{2}}\left(
1-\frac{m_{0}}{r}-\frac{R_{0}r^{2}}{12}+\frac{q^{2}e^{-\gamma }}{\left(
1+f_{R_{0}}\right) r^{2}}-\frac{c}{r^{3w+1} }\right).
    \label{bb6}
\end{equation}

\begin{figure}[ht!]
    \centering
    \includegraphics[width=0.45\linewidth]{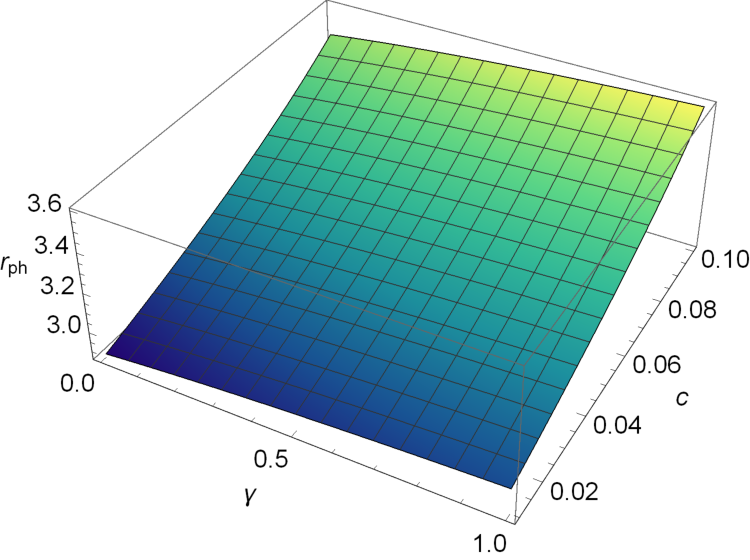}\quad\quad
    \includegraphics[width=0.45\linewidth]{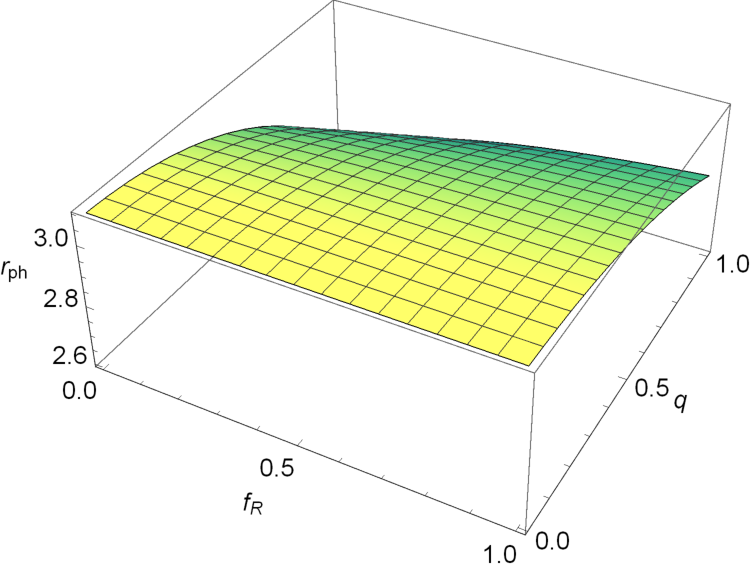}
    \caption{\raggedright Behavior of $r_{\rm ph}$ for numerous values of BH parameters $\gamma$  and $c$ for fixed $f_{R_{0}}$ and $q$ (left), for various values of $f_{R_{0}}$ and $q$ for fixed $\gamma$  and $c$ (right). Here, $m_0=1,R_0=-0.01$ and $w=-2/3$.}
    \label{figa1}
\end{figure}

The unstable circular photon trajectories largely determine the form of the BH’s shadow since incoming photons can either be captured, scattered, or trapped in circular motion near the horizon. The identification of these orbits follows from implementing the following constraints
\begin{equation}
\left. V_{eff}\left( r\right) \right\vert _{r=r_{\rm ph}}= \,\,\,
\left. \frac{d}{dr}V_{eff}\left( r\right) \right\vert _{r=r_{\rm ph}}=0, \,\,\,
\text{and}  \,\, \left. \frac{d^{2}}{dr^{2}}V_{eff}\left( r\right) \right\vert_{r=r_{\rm ph}}<0, \label{con1}
\end{equation}
which shows that the effective potential attains a peak at the radius of the photon sphere $r_{\rm ph}$. Using  Eq. (\ref{con1}), we obtain
\begin{equation}
r_{\rm ph}\,g^{\prime }\left( r_{\rm ph}\right) -2g\left( r_{\rm ph}\right) =0,
\label{psr}
\end{equation}
in terms of $r_{\rm ph}$.  After plugging Eq.~(\ref{m2}) into Eq.~(\ref{psr}), we derive the relation for the photon sphere, which takes the following form
\begin{equation}
r\left(6\,m_0\,-2\,r+3\,c\,(1+w)r^{-3w}\right)(1+f_{R_{0}})-4q^2e^{-\gamma}=0. \label{eps1}
\end{equation}
 When switching off the parameters  $c=0=f_{R_{0}}=0=q$, then  Eq. (\ref{eps1}) reduces to $3\,m_0$. Because of the particular form of the present form of our BH metric, exact analytical formulas for the photon sphere radius can be determined for the different values of the quintessence state parameter $\omega$. This feature is particularly significant because, in higher-order curvature gravity, photon sphere radii are generally determined only by numerical analysis. Therefore, the solution of Eq.~\eqref{eps1} depends on the choice of the quintessence state parameter $w$. Analytically, we obtain the photon sphere for three distinct  selections of the state parameter as follows:
\begin{eqnarray}
&&(i)\,\,\mbox{for}\,\, w=-1/3, \,\,  \Rightarrow r_{\rm ph}=\frac{3m_0+\sqrt{9m_0^2(1+f_{R_{0}})+32q^2(c-1)e^{-\gamma}}}{4(1-c)\sqrt{1+f_{R_{0}}}},
 \\
&&(ii)\,\,\mbox{for}\, \, w=-2/3,\,\,  \Rightarrow
r_{\rm ph}=\frac{2}{3c}+\frac{(1-i\sqrt{3})A_1}{3(2)^{2/3}c\left(4A_2+\sqrt{A_1+A_2^2}\right)^{1/3}}-\frac{(1+i\sqrt{3})\left(4A_2+\sqrt{A_1+A_2^2}\right)^{1/3}}{6(2)^{1/3}c},\\
&&(iii)\,\,\mbox{for}\,\,  w=-1,\,\,   \Rightarrow r_{\rm ph}=\frac{1}{4}\left(3m_0+\sqrt{9m_0^2- \frac{32q^2e^{-\gamma}}{1+f_{R_{0}}}}\right) , \label{bb16c}
\end{eqnarray}
where $A_1=18\,c\,m_0-4$ and  $A_2=16-108\,c\,m_0+\frac{108\,c^2\,q^2e^{-\gamma}}{1+f_{R_{0}}}$.\\ Fig.~\ref{figa1} presents a detailed analysis of the impact of BH parameters on $r_{\rm ph}$ for the special choice $w=-2/3$. The left panel displays the dependence of the photon sphere radius $r_{\rm ph}$   on the parameters $\gamma$ and $c$, with $f_{R_{0}}$ and $q$ held fixed. It demonstrates that $r_{\rm ph}$
  increases monotonically with both $\gamma$ and $c$. In contrast, the right panel illustrates the variation of $r_{\rm ph}$
  with respect to $f_{R_{0}}$ and $q$, while keeping $\gamma$ and $c$ constant. Here, the photon sphere radius exhibits a positive correlation with
$f_{R_{0}}$ but a negative correlation with
$q$. Notably, the parameters
$\gamma$, $c$, and $f_{R_{0}}$ exhibit distinct effects on $r_{\rm ph}$
  compared to the influence of
$q$, highlighting their differing roles in shaping the photon sphere. We note that the photon sphere exhibits similar behavior for different values of the quintessence state parameter $(w = -1/3, -1)$, just as in the case of $w = -2/3$.

 Since the spacetime is not asymptotically flat, the BH shadow must be defined with respect to a static observer located at a finite radius $r=r_{o}$. This formulation ensures that the shadow size remains physically consistent and allows for a systematic comparison between different choices of parameters.
At the photon sphere, the impact parameter is obtained from
\begin{equation}
E^{2}=\frac{L^{2}}{r_{\mathrm{ph}}^{2}}\,g(r_{\mathrm{ph}}),
\end{equation}
which gives
\begin{equation}
b_{\mathrm{ph}}^{2}\equiv\left(\frac{L}{E}\right)^{2}
=\frac{r_{\mathrm{ph}}^{2}}{\left(
1-\frac{m_{0}}{r_{\mathrm{ph}}}-\frac{R_{0}r_{\mathrm{ph}}^{2}}{12}+\frac{q^{2}e^{-\gamma }}{\left(
1+f_{R_{0}}\right) r_{\mathrm{ph}}^{2}}-\frac{c}{r_{\mathrm{ph}}^{3w+1} }\right)}.
\end{equation}

Projecting the photon four-momentum onto the orthonormal tetrad of the static observer, the observable shadow radius on the celestial sky is
\begin{equation}
R_{\mathrm{sh}}
=b_{\mathrm{ph}}\sqrt{g(r_{o})}
=\frac{r_{\mathrm{ph}}}{\sqrt{g(r_{\mathrm{ph}})}}\sqrt{g(r_{o})}.
\end{equation}
In the asymptotically flat limit $g(r_{o})\rightarrow1$, the standard expression for the shadow radius is recovered. A better understanding of the impact of BH parameters can be obtained by tabulating different values of $r_{\rm ph}$ and $R_{\rm s}$ numerically. Tables \ref{taba13} and \ref{taba131} demonstrate how variations in the BH parameters qualitatively influence the $r_{\rm ph}$ and $R_{\rm s}$ for the specific choice of the quintessence parameter $w=-2/3$.

\begin{center}
\begin{tabular}{|c|c c|c c|c c|}
 \hline
 \multicolumn{7}{|c|}{ $c=0.01$}
\\  \hline
 &  \multicolumn{2}{|c|}{$f_{R_0}=0.5$}   & \multicolumn{2}{|c|}{$f_{R_0}=1$}  & \multicolumn{2}{|c|}{$f_{R_0}=1.5$}    \\ \hline $\gamma $  & $r_{\rm ph}$ & $R_{\rm s}$ & $r_{\rm ph}$ & $R_{\rm s}$ & $r_{\rm ph}$ & $R_{\rm s}$ \\ \hline
$0.2$ & $2.90707$ & $3.37396$ & $2.94317$ & $3.42065$ & $2.9644$ & $3.43986$
\\
$0.6$ & $2.95448$ & $3.44458$ & $2.978$ & $3.46568$ & $2.99193$ & $3.47819$
\\
$1$ & $2.98542$ & $3.48145$ & $2.98542$ & $3.48145$ & $3.01011$ & $3.5035$
\\ \hline
 \multicolumn{7}{|c|}{ $c=0.03$}
\\  \hline
$0.2$ & $3.00462$ & $3.1943$ & $3.0419$ & $3.23997$ & $.06385$ & $3.26686$
\\
$0.6$ & $3.0536$ & $3.2543$ & $3.07791$ & $3.2841$ & $3.09232$ & $3.28934$
\\
$1$ & $3.08558$ & $3.29351$ & $3.1016$ & $3.31314$ & $3.11113$ & $3.32483$
\\
 \hline
  \multicolumn{7}{|c|}{ $c=0.05$}
\\  \hline
$0.2$ & $3.11677$ & $2.92335$ & $3.15553$ & $2.95513$ & $3.17836$ & $2.9739$
\\
$0.6$ & $3.1677$ & $2.98057$ & $3.19301$ & $3.0012$ & $3.20801$ & $3.01344$
\\
$1$ & $3.20099$ & $3.01803$ & $3.21768$ & $3.03158$ & $3.23717$ & $3.04744$
\\ \hline
\end{tabular}
\captionof{table}{Numerical values for the $r_{\rm ph}$ and $R_{\rm s}$ with various BH parameters. Here, $m_0=1,q=0.6,R_0=-0.01, r_{o}=10M$ and $w=-2/3$.} \label{taba13}
\end{center}
\begin{center}
\begin{tabular}{|c|c c|c c|c c|}
 \hline
 \multicolumn{7}{|c|}{ $ \gamma=0.6, f_{R_0}=1$}
\\  \hline
 &  \multicolumn{2}{|c|}{$c=0.01$}   & \multicolumn{2}{|c|}{$c=0.03$}  & \multicolumn{2}{|c|}{$c=0.05$}    \\ \hline $q $  & $r_{\rm ph}$ & $R_{\rm s}$ & $r_{\rm ph}$ & $R_{\rm s}$ & $r_{\rm ph}$ & $R_{\rm s}$ \\ \hline
$0.3$ & $3.02959$ & $3.53233$ & $3.1313$ & $3.34957$ & $3.24864$ & $3.07174$
\\
$0.6$ & $2.978$ & $3.46568$ & $3.07791$ & $3.2841$ & $3.19301$ & $3.00904$
\\
$0.9$ & $2.88776$ & $3.34898$ & $2.98469$ & $3.16988$ & $3.09606$ & $2.90012$
\\
 \hline
\end{tabular}
\captionof{table}{Numerical values for $r_{\rm ph}$ and $R_{\rm s}$ with various charge parameter $q$. Here, $m_0=1$, $r_{o}=10M$ and $R_0=-0.01$.} \label{taba131}
\end{center}

\begin{figure}
    \centering
    \includegraphics[scale=0.65]{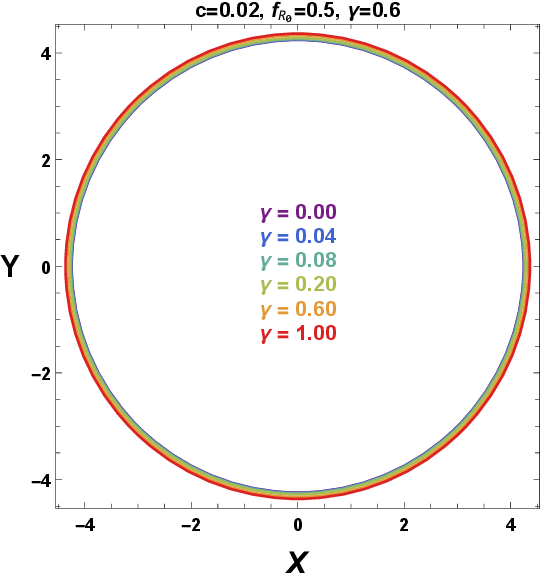}
    \includegraphics[scale=0.65]{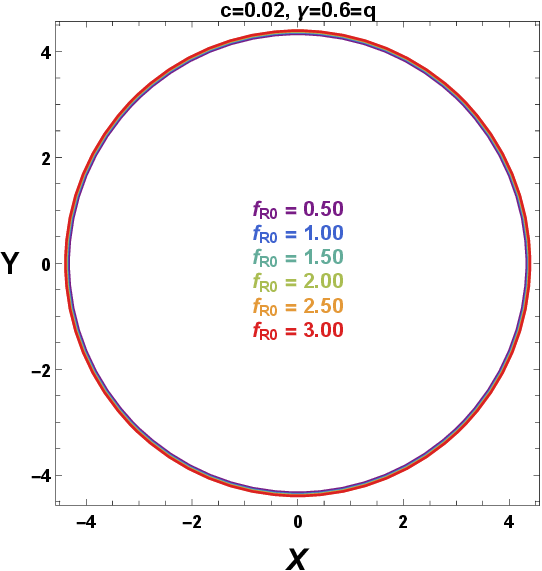}
    \includegraphics[scale=0.65]{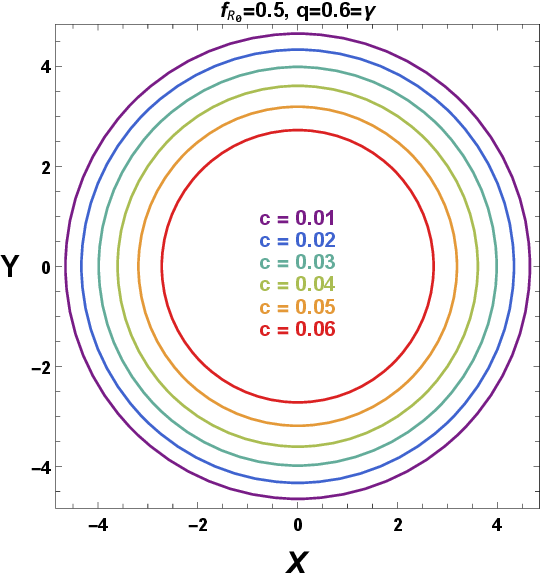}
    \includegraphics[scale=0.65]{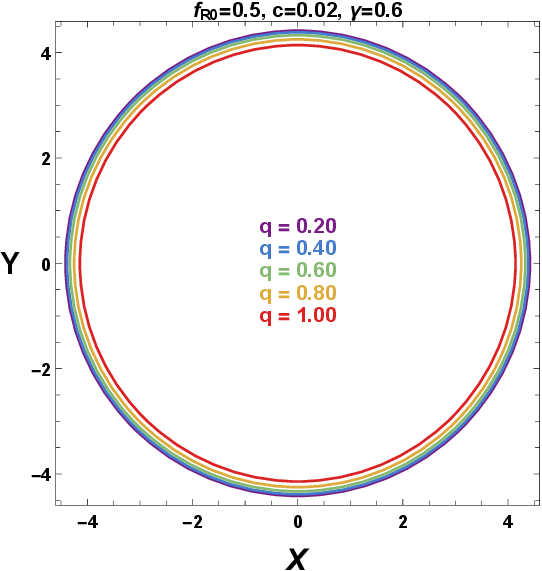}
    \caption{\raggedright BH shadows  for numerous values of $\gamma$ (top left), $f_{R_{0}}$ (top right), $c$ (bottom left) and for various values of $q$ (bottom right). We obtain these graphs by inserting $m_0=1$, $R_0=-0.01$, and $w=-2/3$. }
    \label{ps25}
\end{figure}

In non-asymptotically flat spacetimes, celestial coordinates must be defined with respect to a local observer at a large but finite distance $r_o$. Using the photon's four-momentum, the coordinates are determined by projecting onto the tetrad of the observer \cite{new1,new3,new2}. To characterize the observed shadow of the BH, we utilize the celestial coordinate system $X$ and $Y$
\begin{equation}
X=-\frac{b_{\mathrm{ph}}}{\sqrt{g(r_o)}}\,\csc\theta_o,
\qquad
Y=\pm\sqrt{\frac{\kappa+a^2\cos^2\theta_o-b_{\mathrm{ph}}^2\cot^2\theta_o}{g(r_o)}},
\end{equation}
where $\kappa=K/E^2$, $K$ is Carter constant.\\
 Fig.~\ref{ps25} illustrates the dependence of $R_{\rm s}$ on the $F(R)$-ModMax parameters: the ModMax parameter ($\gamma$), the $F(R)$-gravity parameter ($f_{R_{0}}$), the constant of QF $c$, and the charge ($q$). The shadow radius exhibits a monotonic increase with rising $\gamma,~f_{R_{0}}$, and $c$, whereas it decreases with increasing $q$.

The top left panel of Fig. \ref{ps25}  illustrates the dependence of $R_{\rm s}$ on the ModMax parameter
$\gamma$. The figure demonstrates a monotonic increase in the shadow radius with increasing $\gamma$ (top left panel). This pattern also applies to the parameter $f_{R_{0}}$ (top right panel). In contrast, the left bottom panel shows a monotonic decrease in the shadow size as the parameter $c$ increases.   Similarly, the shadow size exhibits a decreasing behavior with increasing charge parameter
$q$, as shown in the bottom right panel. Our results reveal that quintessence influences the shadow size more strongly than electric charge, highlighting the dominant role of DE in BH optical observables in higher-order curvature gravity.  Table \ref{tab:4} summarizes the analytical expressions of the photon-sphere radius $r_{ph}$ and the corresponding shadow radius $R_s$ for several limiting black-hole configurations within ModMax electrodynamics and its extensions. Each row illustrates how successive physical ingredients, namely $F(R)$ corrections, quintessence fields, and nonlinear electromagnetic effects—modify the location of the photon sphere and the size of the black-hole shadow. The results consistently reduce to the known Reissner–Nordström–(A)dS case in the appropriate parameter limits.
\begin{center}
    \begin{table}[ht!]
\begin{tabular}{|l|l|l|}
\hline
\textbf{ BH solutions } & \textbf{Photon Sphere} & \textbf{Shadow Radius \( R_{\rm s} \)} \\
\hline
\( \text{ $F(R)$-ModMax} \) &
\( r_{\text{ph}}=\frac{1}{2}\left( 3m_0+\sqrt{9m_0^2-\frac{8q^2e^{\gamma}}{1+f_{R_0}}} \right) \) &
\( \displaystyle R_{\rm s} = \frac{r_{\rm ph}}{\sqrt{1-\frac{m_{0}}{r_{\rm ph}}-\frac{R_{0}r^{2}_{ph}}{12}+\frac{q^{2}e^{-\gamma }}{\left(
1+f_{R_{0}}\right) r^{2}_{ph}}}} \)  \\
\hline
\( \text{ ModMax with QF}\) &
\( r_{\text{ph}}=\frac{-6m_0-\sqrt{36m_0^2+32q^2(c-1)e^{-\gamma}}}{4(c-1)} \) &
\( \displaystyle R_{\rm s} = \frac{r_{\rm ph}}{\sqrt{1-\frac{m_{0}}{r_{\rm ph}}-\frac{R_{0}r^{2}_{ph}}{12}+\frac{q^{2}e^{-\gamma }}{ r^{2}_{ph}}-\frac{c}{r_{\rm ph}^{3\omega+1}}}} \)  \\
\hline
\( \text{ ModMax}\) &
\( r_{\text{ph}} = \frac{1}{2}\left( 3m_0+\sqrt{9m_0^2-{8q^2e^{\gamma}}} \right)\) &
\( \displaystyle R_{\rm s} = \frac{r_{\rm ph}}{\sqrt{1-\frac{m_{0}}{r_{\rm ph}}-\frac{R_{0}r^{2}_{ph}}{12}+\frac{q^{2}e^{-\gamma }}{ r^{2}_{ph}}}} \)  \\
\hline

\(\text{  Reissner-Nordstr\"{o}%
m-(A)dS } \) &
\( r_{\text{ph}} =\frac{1}{2}\left( 3m_0+\sqrt{9m_0^2-{8q^2}} \right) \) &
\( \displaystyle R_{\rm s} = \frac{r_{\rm ph}\,g(r_0)}{\sqrt{1-\frac{m_{0}}{r_{\rm ph}}-\frac{R_{0}r^{2}_{ph}}{12}+\frac{q^{2}}{ r^{2}_{ph}}}} \)  \\
\hline
\end{tabular}
\caption{\raggedright Summary of $r_{\rm ph}$ and $R_{\rm s}$ for various parameter assumptions. When $c\neq0$ we use $w=-1/3.$}
\label{tab:4}
\end{table}
\end{center}
Furthermore, this analysis suggests that the conventional BH shadow predicted by GR can be modified when DE, nonlinear electrodynamics, and higher-order curvature corrections act together. Our study of the photon sphere and shadows shows that the derived BH solution exhibits notable optical characteristics shaped by Higher-order curvature gravity ($F(R)$ gravity), ModMax nonlinear electrodynamics, and quintessence. The presence of exact analytical expressions for the photon sphere, together with a consistent incorporation of the finite-distance observers, further enhances the robustness of our results.  These characteristics demonstrate that BH shadows are an effective tool for examining higher-order curvature gravities and DE in astrophysical settings.

\section{Conclusions}\label{sec5}

The present analysis yields an explicit solution to the equations of the gravitational field  corresponding to a novel setup classified by parameters such as mass
$m_0$, charge
$q$, cosmological constant
$R_0=4\,\Lambda$, the ModMax parameter
$\gamma$, the
$F(R)$  parameter $f_{R_0}$, and the QF parameters
$c,w$. In suitable limits, the metric function given in Eq.~(\ref{g(r)F(R)}) recovers established BH solutions, thereby confirming the robustness and consistency of our formulation. This work focuses on examining the influence of QF and
$F(R)-$gravity on the observable astrophysical features around a ModMax BH, emphasizing shadow formation and thermodynamic aspects. For this purpose, the thermodynamic variables of the ModMax BH coupled with a QF in $F(R)-$gravity, such as mass \eqref{m2}, temperature \eqref{T1}, pressure, and volume \eqref{PV}, are analytically obtained in the form of entropy and we also graphically presented their behavior in Figs.~\ref{fig-1a}-\ref{fig-1c}, respectively. Furthermore, we also study the thermal stability of the ModMax BHs with QF in $F(R)$-gravity by deriving the heat capacity and free energies (HFE and GFE), which are associated with the local and global stability of the BHs. We also mention here that, in our analysis, we graphically compare different solutions of the ModMax BHs with QF in GR and in $F(R)$-gravity.

We observed that QF and the $F(R)$-gravity play a significant role in the stability of different types of ModMax BHs. As demonstrated in Figs.~\ref{fig-2}-\ref{fig-4}, the ModMax BH exhibits local thermodynamic instability but remains globally stable across the considered parameter space. The presence of negative heat capacity in certain domains indicates that minor temperature perturbations can escalate unboundedly, suggesting local instability in the system. HFE steadily decreases for small entropy values, indicating that the system naturally transitions toward a lower-energy state. Meanwhile, GFE exhibits an initially decreasing and then increasing behavior, without any swallow-tail behavior, indicating the absence of phase transitions and confirming the system's global thermodynamic stability.

We have further evaluated the sparsity of Hawking radiation and the energy emission rate using the entropy relation. As previously noted, while describing the behavior in Figs.~\ref{fig-55},~\ref{fig-6}, our study evaluated the sparsity $\eta$ and $\varepsilon_{\omega t}$ by employing the temperature given in Eq.~\eqref{T1}. We found that the sparsity parameter increases steeply when entropy is small, then decreases, but as the entropy grows substantially, it increases again gradually for both ModMax BH with QF in $F(R)$-gravity and without $F(R)$-gravity. While in the case of without QF, the ModMax BHs with and without $F(R)$-gravity showed that for high ranges of entropy, $\eta$ becomes constant as illustrated in Fig.~\ref{fig-55}. Similarly, in the energy emission analysis, we found that all ModMax BH models converged to zero for higher entropy values, as discussed in Fig.~\ref{fig-6}. This behavior of the $\varepsilon_{\omega t}$ indicates that the evaporation for smaller BHs is slower.

We analyzed null geodesics to determine $r_{\rm ph}$ and $R_{\rm s}$. Our findings reveal significant influences from both the $F(R)$-ModMax parameters ($\gamma$, $f_{R_0}$, $q$) and the quintessence parameters ($c$, $w$). We calculated analytical relations for both $r_{\rm ph}$ and $R_{\rm s}$.  As the parameters ($\gamma,f_{R_0}$) rise, so do
 the radii of the photon sphere and shadow (Figs.~\ref{figa1} and \ref{ps25}). However, as the parameters $c,q$ increase, the
 shadow size shrinks (Figs.~\ref{figa1} and \ref{ps25}).  The size of the shadow of the  ModMax BHs with QF in $F(R)$-gravity is significantly dependent on the BH parameters (Fig.~\ref{ps25}). For higher values of the parameters ($\gamma,f_{R_0}$), the shadow size grows dramatically. In contrast, as $q$ and $c$ grow, the shadow size decreases.

This study opens several promising avenues for further research. First, extending the stability analysis of the charged BH in $F(R)$-ModMax gravity coupled to a QF to include scalar, gravitational, and electromagnetic perturbations would provide a more comprehensive understanding of the solution's stability. Second, computing the quasinormal modes could yield valuable information about the gravitational-wave signatures of these BHs, potentially enabling observational investigation. Third, studying optical features, such as gravitational lensing, could provide possible observational signatures for analyzing this theory. We intend to address these issues in our future studies.

\section*{Acknowledgments}
{The work of K.B. was partially supported by the JSPS KAKENHI Grant Numbers 24KF0100, 25KF0176, and Competitive Research Funds for Fukushima University Faculty (25RK011). Furthermore, we sincerely acknowledge the anonymous referees for their valuable comments and suggestions.}

\section*{Funding Information}

No Funding Information.

\section*{Conflict of Interest}

Author(s) declares no conflict of interest.

\section*{Data Availability Statement}

No new data were generated or analyzed in this article.


\begin{thebibliography}{999}

\bibitem{nq1}L. Amendola, S. Tsujikawa, DE: Theory and
Observations (Cambridge University Press, Cambridge,
2010).
 
\bibitem{mxxx1} A.~De Felice and S.~Tsujikawa,
Living Rev. Rel. \textbf{13}, 3 (2010).

\bibitem{mxxx2} S.~Capozziello and M.~De Laurentis,
Phys. Rept. \textbf{509}, 167-321 (2011).


\bibitem{mxxx3} 
A.~A.~Starobinsky,
JETP Lett. \textbf{86}, 157-163 (2007).

 \bibitem{mxxx4} Y.~F.~Cai, S.~Capozziello, M.~De Laurentis and E.~N.~Saridakis,
Rept. Prog. Phys. \textbf{79}, no.10, 106901 (2016).

 \bibitem{mxxx5} B.~Li, T.~P.~Sotiriou and J.~D.~Barrow,
Phys. Rev. D \textbf{83}, 064035 (2011).

\bibitem{nq2} T.~Harko, F.~S.~N.~Lobo, S.~Nojiri and S.~D.~Odintsov,
Phys. Rev. D \textbf{84}, 024020 (2011).
[arXiv:1104.2669 [gr-qc]].

\bibitem{mxxx6} P.~Rastall,
Phys. Rev. D \textbf{6}, 3357-3359 (1972).

\bibitem{mxxx7} P.~Rastall,
Can. J. Phys. \textbf{54}, 66-75 (1976).

\bibitem{nq3} P.~Sarmah, A.~De and U.~D.~Goswami,
Phys. Dark Univ. \textbf{40}, 101209 (2023).

\bibitem{nq7} N.~Parbin and U.~D.~Goswami,
Eur. Phys. J. C \textbf{83}, no.5, 411 (2023).

\bibitem{nq6} N.~Parbin and U.~D.~Goswami,
Mod. Phys. Lett. A \textbf{36}, no.37, 2150265 (2021).

\bibitem{nq8} R.~Myrzakulov,
Eur. Phys. J. C \textbf{71}, 1752 (2011).

\bibitem{nq9} N.~Aghanim \textit{et al.} [Planck],
Astron. Astrophys. \textbf{641}, A6 (2020).

\bibitem{nq10} P.~Bull \textit{et al.}
Phys. Dark Univ. \textbf{12}, 56-99 (2016).

\bibitem{mx3} M.~Akbar and R.~G.~Cai,
Phys. Lett. B \textbf{635}, 7-10 (2006).

\bibitem{mx4} G.~Cognola, E.~Elizalde, S.~Nojiri, S.~D.~Odintsov, L.~Sebastiani and S.~Zerbini,
Phys. Rev. D \textbf{77}, 046009 (2008).

\bibitem{mx5} S.~Perlmutter \textit{et al.} [Supernova Cosmology Project],
Astrophys. J. \textbf{517}, 565-586 (1999).
 
 \bibitem{mx6} A.~G.~Riess \textit{et al.} [Supernova Search Team],
Astrophys. J. \textbf{607}, 665-687 (2004).
 
 \bibitem{mx7} S.~Capozziello and A.~Troisi,
Phys. Rev. D \textbf{72}, 044022 (2005).
 
 \bibitem{mx8} S.~Capozziello, A.~Stabile and A.~Troisi,
Phys. Rev. D \textbf{76}, 104019 (2007).
 
 \bibitem{mx9} S.~H.~Hendi, B.~Eslam Panah and S.~M.~Mousavi,
Gen. Rel. Grav. \textbf{44}, 835-853 (2012).

\bibitem{NojiriO2003} S.~Nojiri and S.~D.~Odintsov,
Phys. Rev. D \textbf{68}, 123512 (2003).

\bibitem{NojiriO2011} S.~Nojiri and S.~D.~Odintsov,
Phys. Rept. \textbf{505}, 59-144 (2011).

\bibitem{Mod1} A.~A.~Starobinsky,
Phys. Lett. B \textbf{91}, 99-102 (1980).

\bibitem{Mod3} L.~Amendola and S.~Tsujikawa,
Phys. Lett. B \textbf{660}, 125-132 (2008).


\bibitem{Mod6} S.~Capozziello, E.~Piedipalumbo, C.~Rubano and P.~Scudellaro,
Astron. Astrophys. \textbf{505}, 21-28 (2009).

\bibitem{Mod7} A.~V.~Astashenok, S.~Capozziello and S.~D.~Odintsov,
JCAP \textbf{12}, 040 (2013).

\bibitem{Mod7a} S.~D.~Odintsov and V.~K.~Oikonomou,
Phys. Lett. B \textbf{833}, 137353 (2022).

\bibitem{Mod7b} S.~D.~Odintsov, V.~K.~Oikonomou and G.~S.~Sharov,
Phys. Lett. B \textbf{843}, 137988 (2023).

\bibitem{mxx1} M. Born, and L. Infeld, Foundations of the new field theory, Nature 132 (1933) 1004; 
Proc. R. Soc. Lond. 144 (1934) 425.

\bibitem{mx10} I.~Bandos, K.~Lechner, D.~Sorokin and P.~K.~Townsend,
Phys. Rev. D \textbf{102}, 121703 (2020).

\bibitem{md8} M.~Zhang and J.~Jiang,
Phys. Rev. D \textbf{104}, no.8, 084094 (2021).

\bibitem{md9} I.~Bandos, K.~Lechner, D.~Sorokin and P.~K.~Townsend,
JHEP \textbf{10}, 031 (2021).

\bibitem{md11} S.~I.~Kruglov,
Int. J. Mod. Phys. D \textbf{31}, no.04, 2250025 (2022).

\bibitem{md12} A.~Ali and K.~Saifullah,
Annals Phys. \textbf{437}, 168726 (2022).

\bibitem{md13} A.~Banerjee and A.~Mehra,
Phys. Rev. D \textbf{106}, no.8, 085005 (2022).

\bibitem{md14} I.~Bandos, K.~Lechner, D.~Sorokin and P.~K.~Townsend,
JHEP \textbf{03}, 022 (2021).

\bibitem{md15} J.~Barrientos, A.~Cisterna, D.~Kubiznak and J.~Oliva,
Phys. Lett. B \textbf{834}, 137447 (2022).

\bibitem{mxx5} B.~P.~Kosyakov,
Phys. Lett. B \textbf{810}, 135840 (2020).

\bibitem{mxx6} S.~M.~Kuzenko and E.~S.~N.~Raptakis,
JHEP \textbf{06}, 162 (2024). 

\bibitem{mxx7} K.~Lechner, P.~Marchetti, A.~Sainaghi and D.~P.~Sorokin,
Phys. Rev. D \textbf{106}, no.1, 016009 (2022).

\bibitem{mxx8} B.~P.~Kosyakov,
Phys. Lett. B \textbf{810}, 135840 (2020). 


\bibitem{mxx10} S.~I.~Kruglov,
Int. J. Mod. Phys. D \textbf{31}, no.04, 2250025 (2022).

\bibitem{mxx12} Y.~Sekhmani, S.~K.~Maurya, M.~K.~Jasim, A.~Al-Badawi and J.~Rayimbaev,
Phys. Dark Univ. \textbf{46}, 101701 (2024).

\bibitem{mxx13} R.~C.~Pantig, L.~Mastrototaro, G.~Lambiase and A.~{\"O}vg{\"u}n,
Eur. Phys. J. C \textbf{82}, no.12, 1155 (2022).

\bibitem{mxx14}  A.~Al-Badawi, Y.~Sekhmani and K.~Boshkayev,
Phys. Dark Univ. \textbf{48}, 101865 (2025).

\bibitem{mxx15} Z. Amirabi, Ann. Phys. 443 (2022) 168990.

\bibitem{mxx16} B.~Eslam Panah,
PTEP \textbf{2024}, no.2, 023E01 (2024).

\bibitem{EslamPanah:2024lbk}
B.~Eslam Panah,
[arXiv:2410.16346 [gr-qc]].

\bibitem{Siahaan:2024ioa}
H.~M.~Siahaan,
Phys. Lett. B \textbf{865}, 139479 (2025).

\bibitem{Barrientos:2024umq}
J.~Barrientos, A.~Cisterna, M.~Hassaine and K.~Pallikaris,
Phys. Lett. B \textbf{860}, 139214 (2025).

\bibitem{Guzman-Herrera:2024fkg}
E.~Guzman-Herrera, A.~Montiel and N.~Breton,
JCAP \textbf{11}, 002 (2024).

\bibitem{EslamPanah:2024fls}
B.~Eslam Panah, B.~Hazarika and P.~Phukon,
PTEP \textbf{2024}, no.8, 083E02 (2024).

\bibitem{Shahzad:2024ljt}
M.~R.~Shahzad, G.~Abbas, H.~Rehman and W.~X.~Ma,
Eur. Phys. J. C \textbf{84}, no.5, 461 (2024).

\bibitem{Russo:2024xnh}
J.~G.~Russo and P.~K.~Townsend,
JHEP \textbf{06}, 191 (2024).

\bibitem{Ayon-Beato:2024vph}
E.~Ay{\'o}n-Beato, D.~Flores-Alfonso and M.~Hassaine,
Phys. Rev. D \textbf{110}, no.6, 064027 (2024).

\bibitem{Guzman-Herrera:2023zsv}
E.~Guzman-Herrera and N.~Breton,
JCAP \textbf{01}, 041 (2024).

\bibitem{Bakhtiarizadeh:2023mhk}
H.~R.~Bakhtiarizadeh and H.~Golchin,
JCAP \textbf{01}, 061 (2024).

\bibitem{Rathi:2023vhw}
H.~Rathi and D.~Roychowdhury,
JHEP \textbf{07}, 026 (2023).


\bibitem{Flores-Alfonso:2020nnd}
D.~Flores-Alfonso, R.~Linares and M.~Maceda,
JHEP \textbf{09}, 104 (2021).

\bibitem{BallonBordo:2020jtw}
A.~Ballon Bordo, D.~Kubiz{\v{n}}{\'a}k and T.~R.~Perche,
Phys. Lett. B \textbf{817}, 136312 (2021).

\bibitem{Flores-Alfonso:2020euz}
D.~Flores-Alfonso, B.~A.~Gonz{\'a}lez-Morales, R.~Linares and M.~Maceda,
Phys. Lett. B \textbf{812}, 136011 (2021).

\bibitem{Sekhmani:2025epe}
Y.~Sekhmani, S.~K.~Maurya, J.~Rayimbaev, M.~Altanji, I.~Ibragimov and S.~Muminov,
Phys. Dark Univ. \textbf{50}, 102079 (2025).

\bibitem{Anand:2025iwc}
A.~Anand and R.~Kumar,
Phys. Lett. B \textbf{868}, 139807 (2025).

\bibitem{Sekhmani:2025jbl}
Y.~Sekhmani, A.~Baruah, S.~K.~Maurya, J.~Rayimbaev, M.~Altanji, I.~Ibragimov and S.~Muminov,
Phys. Dark Univ. \textbf{50}, 102157 (2025).

\bibitem{EslamPanah:2025bfh}
B.~Eslam Panah,
Phys. Lett. B \textbf{868}, 139711 (2025).

\bibitem{Baptista:2025ogh}
A.~Q.~Baptista and M.~L.~Pe{\~n}afiel,
Phys. Rev. D \textbf{112}, no.2, 024037 (2025).

\bibitem{Diaz:2025zuc}
J.~M.~Diaz and M.~E.~Rubio,
Phys. Rev. D \textbf{111}, no.12, 124030 (2025).

\bibitem{Barrientos:2025rde}
J.~Barrientos, N.~C{\'a}ceres, F.~Diaz and U.~Hernandez-Vera,
Phys. Rev. D \textbf{112}, no.8, 086018 (2025).

\bibitem{Yasir:2025xgh}
M.~Yasir, F.~Mushtaq, F.~Javed, M.~Alosaimi and R.~M.~Zulqarnain,
Phys. Dark Univ. \textbf{48}, 101929 (2025).

\bibitem{Heidari:2025llu}
N.~Heidari and B.~Eslam Panah,
Phys. Lett. B \textbf{866}, 139530 (2025).

\bibitem{NooriGashti:2025jmi}
S.~Noori Gashti, M.~A.~S.~Afshar, M.~R.~Alipour, {\.I}.~Sakall{\i}, B.~Pourhassan and J.~Sadeghi,
Eur. Phys. J. C \textbf{85}, no.10, 1144 (2025).

\bibitem{Barbosa:2025smt}
S.~Barbosa, S.~Fichet and L.~de Souza,
JHEP \textbf{10}, 145 (2025).



\bibitem{Hale:2025veb}
T.~Hale, D.~Kubiz{\v{n}}{\'a}k, J.~Men{\v{s}}{\'\i}kov{\'a}, R.~B.~Mann and J.~Yang,
Phys. Rev. D \textbf{111}, no.10, 104004 (2025).

\bibitem{Bokulic:2025usc}
A.~Bokuli{\'c} and C.~A.~R.~Herdeiro,
Phys. Rev. D \textbf{111}, no.6, 064046 (2025).

\bibitem{EslamPanah:2024gxx}
B.~Eslam Panah and N.~Heidari,
JHEAp \textbf{45}, 181-193 (2025).

\bibitem{Kurbonov:2026oem}
N.~Kurbonov, S.~Rizaev, M.~Kurbanova, J.~Rayimbaev, M.~Zahid and M.~Akhmedov,
Phys. Dark Univ. \textbf{52}, 102300 (2026).

\bibitem{EslamPanah:2025oqy}
B.~Eslam Panah, B.~Hamil and M.~E.~Rodrigues,
Eur. Phys. J. C \textbf{86}, no.1, 81 (2026).

\bibitem{Alloqulov:2025dqi}
M.~Alloqulov, S.~Shaymatov, B.~Ahmedov and T.~Zhu,
Eur. Phys. J. C \textbf{86}, no.3, 259 (2026).

\bibitem{Sekhmani:2025gvv}
Y.~Sekhmani, S.~K.~Maurya, J.~Rayimbaev, M.~Altanji, I.~Ibragimov and S.~Muminov,
Phys. Dark Univ. \textbf{50}, 102116 (2025).

\bibitem{Jafarzade:2024zqq}
K.~Jafarzade, Z.~Bazyar and M.~Jamil,
Phys. Lett. B \textbf{864}, 139390 (2025).


\bibitem{isz16} R.~R.~Caldwell, R.~Dave and P.~J.~Steinhardt,
Phys. Rev. Lett. \textbf{80}, 1582-1585 (1998).

\bibitem{isz17} P.~J.~Steinhardt, L.~M.~Wang and I.~Zlatev,
Phys. Rev. D \textbf{59}, 123504 (1999)

\bibitem{isz18} V.~V.~Kiselev,
Class. Quant. Grav. \textbf{20}, 1187-1198 (2003).

\bibitem{isz19} S.~Chen, B.~Wang and R.~Su,
Phys. Rev. D \textbf{77}, 124011 (2008).

\bibitem{isz20} S.~Fernando,
Gen. Rel. Grav. \textbf{44}, 1857-1879 (2012).

\bibitem{isz21} B.~Malakolkalami and K.~Ghaderi,
Astrophys. Space Sci. \textbf{357}, no.2, 112 (2015).

\bibitem{1}
J.~D.~Bekenstein, 
In Jacob Bekenstein: the conservative revolutionary (pp. 303-306)  (2020).
\bibitem{2}
J.~D.~Bekenstein,
Phys. Rev. D \textbf{7}, 2333-2346 (1973).
\bibitem{3}
D.~Kubiznak, R.~B.~Mann and M.~Teo,
Class. Quant. Grav. \textbf{34}, no.6, 063001 (2017).
\bibitem{4}
J.~Sadeghi, S.~Noori Gashti and E.~Naghd Mezerji,
Phys. Dark Univ. \textbf{30}, 100626 (2020).
\bibitem{5}
V.~V.~Kiselev,
Class. Quant. Grav. \textbf{20}, 1187-1198 (2003).
\bibitem{6}
J.~Sadeghi, M.~Shokri, S.~Gashti Noori and M.~R.~Alipour,
Gen. Rel. Grav. \textbf{54}, no.10, 129 (2022).
\bibitem{7}
Y.~Heydarzade, M.~Misyura and V.~Vertogradov,
Phys. Rev. D \textbf{108}, no.4, 044073 (2023).
\bibitem{8}
J.~Sadeghi, M.~Shokri, M.~R.~Alipour and S.~Noori Gashti,
Chin. Phys. C \textbf{47}, no.1, 015103 (2023).
\bibitem{9}
J.~Rayimbaev, B.~Majeed, M.~Jamil, K.~Jusufi and A.~Wang,
Phys. Dark Univ. \textbf{35}, 100930 (2022).
\bibitem{10}
Y.~Sekhmani, G.~G.~Luciano, J.~Rayimbaev, M.~K.~Jasim, A.~Al-Badawi and S.~K.~Maurya,
Phys. Dark Univ. \textbf{46}, 101567 (2024).
\bibitem{11}
Y.~Sekhmani, D.~J.~Gogoi, M.~Baouahi and I.~Dahiri,
Phys. Scripta \textbf{98}, no.10, 105014 (2023).
\bibitem{12}
S.~W.~Hawking and D.~N.~Page,
Commun. Math. Phys. \textbf{87}, 577 (1983).
\bibitem{13}
J.~Sadeghi, M.~R.~Alipour, M.~A.~S.~Afshar and S.~Noori Gashti,
Gen. Rel. Grav. \textbf{56}, no.8, 93 (2024).
\bibitem{14}
J.~Sadeghi, M.~A.~S.~Afshar, M.~R.~Alipour and S.~Noori Gashti,
Phys. Dark Univ. \textbf{47}, 101780 (2025).
\bibitem{15}
Y.~Sekhmani, D.~J.~Gogoi, R.~Myrzakulov and J.~Rayimbaev,
Commun. Theor. Phys. \textbf{76}, no.4, 045403 (2024).
\bibitem{16}
Sekhmani, Y., et al. "Y.~Sekhmani, J.~Rayimbaev, G.~G.~Luciano, R.~Myrzakulov and D.~J.~Gogoi,
Eur. Phys. J. C \textbf{84}, no.3, 227 (2024).
\bibitem{17}
M.~A.~S.~Afshar, S.~Noori Gashti, M.~R.~Alipour and J.~Sadeghi,
Eur. Phys. J. C \textbf{85}, no.9, 939 (2025).

\bibitem{Al-Badawi:2024mco}
A.~Al-Badawi and A.~Jawad,
Eur. Phys. J. C \textbf{84}, no.2, 115 (2024).

\bibitem{Weinhold:1975xej}
F.~Weinhold,
J. Chem. Phys. \textbf{63}, no.6, 2479 (1975).

\bibitem{Ruppeiner:2008kd}
G.~Ruppeiner,
Phys. Rev. D \textbf{78}, 024016 (2008).

\bibitem{Ruppeiner:1995zz}
G.~Ruppeiner,
Rev. Mod. Phys. \textbf{67}, 605-659 (1995).

\bibitem{Sarkar:2006tg}
T.~Sarkar, G.~Sengupta and B.~Nath Tiwari,
JHEP \textbf{11}, 015 (2006).

\bibitem{Quevedo:2008ry}
H.~Quevedo and A.~Sanchez,
Phys. Rev. D \textbf{79}, 024012 (2009).


\bibitem{Akbar:2011qw}
M.~Akbar, H.~Quevedo, K.~Saifullah, A.~Sanchez and S.~Taj,
Phys. Rev. D \textbf{83}, 084031 (2011).

\bibitem{Soroushfar:2020wch}
S.~Soroushfar and S.~Upadhyay,
Phys. Lett. B \textbf{804}, 135360 (2020).

\bibitem{Hendi:2015cka}
S.~H.~Hendi and R.~Naderi,
Phys. Rev. D \textbf{91}, no.2, 024007 (2015).

\bibitem{Hendi:2015xya}
S.~H.~Hendi, A.~Sheykhi, S.~Panahiyan and B.~Eslam Panah,
Phys. Rev. D \textbf{92}, no.6, 064028 (2015).

\bibitem{Banerjee:2016nse}
R.~Banerjee, B.~R.~Majhi and S.~Samanta,
Phys. Lett. B \textbf{767}, 25-28 (2017).

\bibitem{Bhattacharya:2019qxe}
K.~Bhattacharya and B.~R.~Majhi,
Phys. Lett. B \textbf{802}, 135224 (2020).

\bibitem{NaveenaKumara:2020biu}
A.~Naveena Kumara, C.~L.~Ahmed Rizwan, K.~Hegde, M.~S.~Ali and K.~M.~Ajith,
Phys. Rev. D \textbf{103}, no.4, 044025 (2021).


\bibitem{Bardeen:1973gs}
J.~M.~Bardeen, B.~Carter and S.~W.~Hawking,
Commun. Math. Phys. \textbf{31}, 161-170 (1973).

\bibitem{Hawking:1974rv}
S.~W.~Hawking,
Nature \textbf{248}, 30-31 (1974).

\bibitem{Strominger:1997eq}
A.~Strominger,
JHEP \textbf{02}, 009 (1998).

\bibitem{Hendi:2020mhg}
S.~H.~Hendi, F.~Azari, E.~Rahimi, M.~Elahi, Z.~Owjifard and Z.~Armanfard,
Annalen Phys. \textbf{532}, no.10, 2000162 (2020).

\bibitem{Nashed:2025ebr}
G.~G.~L.~Nashed, U.~Zafar and K.~Bamba,
Phys. Dark Univ. \textbf{50}, 102061 (2025).

\bibitem{Page:1976ki}
D.~N.~Page,
Phys. Rev. D \textbf{14}, 3260-3273 (1976).

\bibitem{Gray:2015pma}
F.~Gray, S.~Schuster, A.~Van-Brunt and M.~Visser,
Class. Quant. Grav. \textbf{33}, no.11, 115003 (2016).

\bibitem{mxx20} B.~P.~Kosyakov,
Phys. Lett. B \textbf{810}, 135840 (2020).

\bibitem{Ghosh:2017cuq} D.~Kubiznak, T.~Tahamtan and O.~Svitek,
Phys. Rev. D \textbf{105}, no.10, 104064 (2022).

\bibitem{Ghosh:2016cuq}Ghosh, S.G.  Eur. Phys. J. C 76, 222 (2016).

\bibitem{Ashtekar:1984zz}
A.~Ashtekar and A.~Magnon,
Class. Quant. Grav. \textbf{1}, L39-L44 (1984).

\bibitem{Wald:1993nt}
R.~M.~Wald,
Phys. Rev. D \textbf{48}, no.8, R3427-R3431 (1993).

\bibitem{Momeni:2025rgk}
D.~Momeni and R.~Myrzakulov,
Int. J. Theor. Phys. \textbf{64}, no.10, 268 (2025).

 \bibitem{Cognola2005} G.~Cognola, E.~Elizalde, S.~Nojiri, S.~D.~Odintsov and S.~Zerbini,
JCAP \textbf{02}, 010 (2005).

\bibitem{Zafar:2025nho}
U.~Zafar, K.~Bamba, A.~Jawad, T.~Rasheed and S.~Shaymatov,
PTEP \textbf{2026}, 023.

\bibitem{Jawad:2023ypn}
A.~Jawad and U.~Zafar,
Nucl. Phys. B \textbf{992}, 116231 (2023)

\bibitem{Kastor:2009wy}
D.~Kastor, S.~Ray and J.~Traschen,
Class. Quant. Grav. \textbf{26}, 195011 (2009).

\bibitem{Rani:2025esb}
S.~Rani, A.~Jawad, M.~Heydari-Fard and U.~Zafar,
Eur. Phys. J. C \textbf{85}, no.6, 677 (2025).

\bibitem{rodrigues2022bardeen}
M.~E.~Rodrigues, M.~V.~de S.~Silva and H.~A.~Vieira,
Phys. Rev. D \textbf{105}, no.8, 084043 (2022).
\bibitem{ruppeiner2014thermodynamic}
G.~Ruppeiner,
Springer Proc. Phys. \textbf{153}, 179-203 (2014).

\bibitem{davies1978thermodynamics}
P. C. W. Davies, Phys. Rept. \textbf{41}, 1313 (1978).

\bibitem{Tariq:2025wiy}
H.~Tariq, U.~Zafar, S.~Chaudhary, K.~Bamba, A.~Jawad and S.~Shaymatov,
Nucl. Phys. B \textbf{1016}, 116906 (2025)

\bibitem{caneva2021helmholtz}
K. L. Caneva, Helmholtz, MIT Press, (2021).

\bibitem{paul2024thermodynamics}
P.~Paul and S.~I.~Kruglov,
Indian J. Phys. \textbf{98}, no.4, 1201-1210 (2024).
\bibitem{simovic2024euclidean}
F.~Simovic and I.~Soranidis,
Phys. Rev. D \textbf{109}, no.4, 044029 (2024).

\bibitem{ali2019thermodynamics}
M.~S.~Ali and S.~G.~Ghosh,
Phys. Rev. D \textbf{99}, no.2, 024015 (2019).

\bibitem{kubizvnak2012p}
D.~Kubiznak and R.~B.~Mann,
JHEP \textbf{07}, 033 (2012).

\bibitem{deng2018thermodynamics}
G.~M.~Deng, J.~Fan, X.~Li and Y.~C.~Huang,
Int. J. Mod. Phys. A \textbf{33}, no.03, 1850022 (2018).

\bibitem{Mushtaq:2025ksk}
F.~Mushtaq, A.~Jawad, X.~Tiecheng, M.~M.~Alam and S.~Shaymatov,
Phys. Dark Univ. \textbf{48}, 101872 (2025).


\bibitem{AA0} A.~Al-Badawi, F.~Ahmed and {\.I}.~Sakall{\i},
Phys. Dark Univ. \textbf{52}, 102245 (2026). 

\bibitem{AA1} A.~Al-Badawi, F.~Ahmed and I.~Sakall{\i},
Commun. Theor. Phys. \textbf{78}, no.2, 025401 (2026).
\bibitem{AA2} A.~Al-Badawi, F.~Ahmed and {\.I}.~Sakall{\i},
Eur. Phys. J. C \textbf{85}, no.6, 660 (2025).

\bibitem{AA3} F.~Ahmed, A.~Al-Badawi and {\.I}.~Sakall{\i},
Nucl. Phys. B \textbf{1017}, 116951 (2025). 

\bibitem{sec3is07} F.~Ahmed, A.~Al-Badawi, {\.I}.~Sakall{\i} and A.~Bouzenadad,
Nucl. Phys. B \textbf{1011}, 116806 (2025).

\bibitem{bozza_gravitational_2010}V. Bozza, Gen. Rel. Grav. 42:2269-2300,(2010).

\bibitem{Carter1968} B.~Carter,
Phys. Rev. \textbf{174}, 1559-1571 (1968).

\bibitem{mxx30} J. W. Gibbs, “On the equilibrium of heterogeneous substances,” American journal of science, vol. 3,
 no. 96, pp. 441–458, (1878).

\bibitem{mxx31} A. F. Tuck, “Gibbs free energy and reaction rate acceleration in and on microdroplets,” Entropy, vol. 21,
 no. 11, p. 1044, (2019).

\bibitem{mxx32} M.~Chabab, A.~El Batoul, I.~El-ilali, A.~Lahbas and M.~Oulne,
Eur. Phys. J. Plus \textbf{135}, 201 (2020).

\bibitem{new1}V. Perlick, O. Tsupko, and G. S. Bisnovatyi-Kogan, 
 Phys. Rev. D 92, 104031 (2015).
 \bibitem{new3} A.~Grenzebach, V.~Perlick and C.~L{\"a}mmerzahl,
Phys. Rev. D \textbf{89}, no.12, 124004 (2014). 

\bibitem{new2} A. Grenzebach, The Shadow of Black Holes,  Cham: Springer International Publishing, 2016.

\end{thebibliography}
\end{document}